\documentclass[%
 reprint,
 preprintnumbers,
superscriptaddress,
twocolumn,
 amsmath,amssymb,
 aps,
 prd,
,fleqn,
]{revtex4-1}

\usepackage{float}
\usepackage{graphicx}
\usepackage{dcolumn}
\usepackage{bm}
\usepackage{booktabs}
\usepackage{xcolor}
\usepackage{verbatim} 
\usepackage{eso-pic}
\usepackage{chngcntr}

\newcommand{\beq}{\begin{equation}}
\newcommand{\eeq}{\end{equation}}

\newcommand{\Halofit}{{\texttt{Halofit}}}

\usepackage[bottom]{footmisc}

\newcommand\be{\begin{equation}}
\newcommand\ee{\end{equation}}
\def\bea{\begin{eqnarray}}
\def\eea{\end{eqnarray}}

\newcommand\mnras{MNRAS}             
\newcommand\apjl{ApJ}                
\newcommand\apjs{ApJS}               

\begin{document}

\preprint{DES-2017-0225}
\preprint{FERMILAB-PUB-17-280-AE}

\title{Dark Energy Survey Year 1 Results: Galaxy clustering for combined probes}



\author{J.~Elvin-Poole}
\affiliation{Jodrell Bank Center for Astrophysics, School of Physics and Astronomy, University of Manchester, Oxford Road, Manchester, M13 9PL, UK}
\author{M.~Crocce}
\affiliation{Institute of Space Sciences, IEEC-CSIC, Campus UAB, Carrer de Can Magrans, s/n,  08193 Barcelona, Spain}
\author{A.~J.~Ross}
\affiliation{Center for Cosmology and Astro-Particle Physics, The Ohio State University, Columbus, OH 43210, USA}


\author{T.~Giannantonio}
\affiliation{Institute of Astronomy, University of Cambridge, Madingley Road, Cambridge CB3 0HA, UK}
\affiliation{Kavli Institute for Cosmology, University of Cambridge, Madingley Road, Cambridge CB3 0HA, UK}
\affiliation{Universit\"ats-Sternwarte, Fakult\"at f\"ur Physik, Ludwig-Maximilians Universit\"at M\"unchen, Scheinerstr. 1, 81679 M\"unchen, Germany}
\author{E.~Rozo}
\affiliation{Department of Physics, University of Arizona, Tucson, AZ 85721, USA}
\author{E.~S.~Rykoff}
\affiliation{SLAC National Accelerator Laboratory, Menlo Park, CA 94025, USA}
\affiliation{Kavli Institute for Particle Astrophysics \& Cosmology, P. O. Box 2450, Stanford University, Stanford, CA 94305, USA}


\author{S.~Avila}
\affiliation{Centro de Investigaciones Energ\'eticas, Medioambientales y Tecnol\'ogicas (CIEMAT), Madrid, Spain}
\affiliation{Institute of Cosmology \& Gravitation, University of Portsmouth, Portsmouth, PO1 3FX, UK}
\author{N.~Banik}
\affiliation{Fermi National Accelerator Laboratory, P. O. Box 500, Batavia, IL 60510, USA}
\affiliation{Department of Physics, University of Florida, Gainesville, Florida 32611, USA}
\author{J.~Blazek}
\affiliation{Institute of Physics, Laboratory of Astrophysics, \'Ecole Polytechnique F\'ed\'erale de Lausanne (EPFL), Observatoire de Sauverny, 1290 Versoix, Switzerland}
\affiliation{Center for Cosmology and Astro-Particle Physics, The Ohio State University, Columbus, OH 43210, USA}
\author{S.~L.~Bridle}
\affiliation{Jodrell Bank Center for Astrophysics, School of Physics and Astronomy, University of Manchester, Oxford Road, Manchester, M13 9PL, UK}
% ...

\collaboration{DES Collaboration}

\noaffiliation

\date{\today}
             
\label{firstpage}            

\begin{abstract}
We measure the clustering of DES Year 1 galaxies that are intended to be
combined with weak lensing samples in order to produce precise cosmological constraints from the joint analysis of large-scale structure and lensing correlations. Two-point correlation functions are measured for a sample of $6.6 \times 10^{5}$ luminous red galaxies selected using the \textsc{redMaGiC} algorithm over an area of $1321$ square degrees, in the redshift range $0.15 < z < 0.9$, split into five tomographic redshift bins. The sample has a mean redshift uncertainty of $\sigma_{z}/(1+z) = 0.017$. We quantify and correct spurious correlations induced by spatially variable survey properties, testing their impact on the clustering measurements and covariance. We demonstrate the sample's robustness by testing for stellar contamination, for potential biases that could arise from the systematic correction, and for the consistency between the two-point auto- and cross-correlation functions. We show that the corrections we apply have a significant impact on the resultant measurement of cosmological parameters, but that the results are robust against arbitrary choices in the correction method. We find the linear galaxy bias in each redshift bin in a fiducial cosmology to be 
${b(\sigma_{8}/0.81)\mid_{z=0.24}=1.40 \pm 0.07}$,  ${b(\sigma_{8}/0.81)\mid_{z=0.38}=1.60 \pm 0.05}$,  ${b(\sigma_{8}/0.81)\mid_{z=0.53}=1.60 \pm 0.04}$ for galaxies with luminosities $L/L_*>$$0.5$,  ${b(\sigma_{8}/0.81)\mid_{z=0.68}=1.93 \pm 0.04}$ for $L/L_*>$$1$ and ${b(\sigma_{8}/0.81)\mid_{z=0.83}=1.98 \pm 0.07}$ for $L/L_*$$>1.5$, 
broadly consistent with expectations for the redshift and luminosity dependence of the bias of red galaxies. We show these measurements to be consistent with the linear bias obtained from tangential shear measurements.   
\end{abstract}

\pacs{Valid PACS appear here}

\maketitle


\section{Introduction}
\label{sec:intro}

Galaxies are a biased tracer of the matter density field. In the standard `halo model' paradigm, they form in collapsed over-densities (dark matter halos; \cite{WhiteRees78}), and the mass of the halo they reside in is known to correlate with the luminosity and color of the galaxy, with more luminous and redder galaxies strongly correlated with higher mass. Therefore, the galaxy `bias' depends strongly on the particular sample being studied. Thus, in cosmological studies the galaxy bias is often treated as a nuisance parameter --- one that is degenerate with the amplitude of the clustering of matter. See, e.g., \cite{DesjacquesBiasRev} and references therein.

The degeneracy can be broken with additional observables. This includes the weak gravitational lensing `shear' field, which is induced by the matter density field. Correlation between the galaxies and the shear field (\cite{Tyson84}; often referred to as `galaxy-galaxy lensing') contains one factor of the galaxy bias and two factors of the matter field. The galaxy auto-correlation contains two factors of the galaxy bias and again two factors of the matter field. Thus, the combination of the two measurements can break the degeneracy between the two quantities, and it is a sensitive probe of the late-time matter field (see, e.g., \cite{Mandelbaum13,Kwan17DES})

The auto-correlation of the shear field alone includes no factors of the galaxy bias and can thus be used directly as a probe of the matter field. However, its sensitivity to many systematic uncertainties related to the estimation of the shear field differs from that of the galaxy-galaxy lensing signal. As shown by \cite{Hu:2003pt,Bernstein:2008aq,Joachimi:2009ez,Samuroff17}, the impact of such systematic uncertainties can be mitigated by combining the shear auto-correlation measurements with those of galaxy clustering and galaxy-galaxy lensing. Thus there is substantial gain to be obtained from a combined analysis.

Such a combined analysis is performed with the Dark Energy Survey (DES\footnote{http://www.darkenergysurvey.org/}; \cite{Abbott:2005bi}) Year-1 (Y1) data (\cite{DESY1cos}; hereafter Y1COSMO). DES is an imaging survey currently amassing data over a 5000 deg$^2$ footprint in five passbands ($grizY$). When completed, it will have mapped 300 million galaxies and tens of thousands of galaxy clusters.

In this work, we study the clustering of red sequence galaxies selected from
DES Y1 data using the \textsc{redMaGiC} \citep{Rozo15} algorithm, chosen for its small redshift uncertainty. We study the same sample used to obtain cosmological results in the Y1COSMO combined analysis. In particular, we study the large-scale clustering amplitude and its sensitivity to observational systematics. Following previous studies \cite{Ross11,Ho12,Leistedt13,Crocce16}, we use angular maps to track the observing conditions of the Y1 data in order to identify and correct for spurious fluctuations in the galaxy density field. We further determine the effect the corrections have on the covariance matrix of the angular auto-correlation of the galaxies. We present robust measurements of the clustering amplitude of \textsc{redMaGiC} galaxies as a function of redshift and luminosity, thus gaining insight into the physical nature of these galaxies and how they compare to other red galaxy samples. The results of this paper are then used for the joint DES cosmological analysis presented in Y1COSMO.

This outline of this paper is: we summarize in Section \ref{sec:theory} the model we use to describe our galaxy clustering measurements; we present in Section \ref{sec:data} the DES data we use; Section \ref{sec:methods} presents how we measure clustering statistics and estimate their covariance; Section \ref{sec:systematics} summarizes the results of our observational systematic tests. We present our primary results with galaxy bias measurements in Section \ref{sec:galaxybias} and a demonstration of their robustness in Section \ref{sec:robust} before concluding in Section \ref{sec:conclusions}. 

In order to avoid confirmation bias, we have performed this analysis ``blind'': we did not measure parameter constraints or plot the correlation function measured from the data on the same axis as any theoretical prediction or simulated clustering measurement until the sample and all measurements in Y1COSMO were finalized.

Unless otherwise noted, we use a fiducial $\Lambda {\rm CDM}$ cosmology, fixing cosmological parameters at $\Omega_m = 0.295$, $A_{s} = 2.260574 \times 10^{-9}$, $\Omega_b = 0.0468$, $h = 0.6881$, $n_s = 0.9676$. This is consistent with the latest cosmological data from the Planck mission \citep{Planck15cosmo} and is used as the fiducial cosmology for all the DES Y1 analyses used in Y1COSMO. We use this cosmology to generate Gaussian mocks in Section \ref{sec:systematics} for systematics testing. 

After un-blinding, we re-measure the galaxy bias, fixing the cosmological parameters at the mean of the DES Y1COSMO posterior, $\Omega_m = 0.276$, $A_{s} = 2.818378 \times 10^{-9}$, $\Omega_b = 0.0531$, $h = 0.7506$, $n_s = 0.9939$, $\Omega_{\nu} = 0.00553$ (note that we show these values at a greater precision than can be measured by DES). This cosmology was used for all bias measurements in Section \ref{sec:galaxybias}. 

\section {Theory}
\label{sec:theory}

Throughout this paper we model \textsc{redMaGiC} clustering measurements assuming a local, linear galaxy bias model \citep{FryGaztanaga1993}, where the galaxy and matter density fluctuations are related by $\delta_g({\bf{x}})=b\delta_m(\bf{x})$, with density fluctuations  defined by $\delta \equiv (n({\bf{x}})-{\bar{n}})/{\bar{n}}$. The validity of this assumption over the scales considered here is provided in \cite{MPPmethodology} and shown in simulations in \cite{simspaper}. 

The galaxy clustering model used in this paper matches that used in Y1COSMO. This model includes 3 neutrinos of degenerate mass. 

We consider multiple galaxy redshift bins $i$, each characterized by a 
\textsc{redMaGiC} galaxy redshift distribution $n^i_g(z)$, normalized to unity in redshift, and a galaxy bias $b^i$ which is assumed to be constant across the redshift bin range. Under the Limber \citep{LimberApproximation} and flat-sky approximation the theoretical prediction for the galaxy correlation function $w(\theta)$ in a given bin is, 
\begin{eqnarray}
\!\!\!\!\!\!w^i(\theta) &=&\left(b^i\right)^2\int \frac{dl \; l}{2\pi} J_0(l\theta)\,\int d\chi
 \nonumber \\ 
 &\times&\frac{ \left[n_{\rm g}^i(z(\chi)) dz/d\chi \right]^2}{\chi^2} P_{\mathrm{NL}}\left(\frac{l+1/2}{\chi},z(\chi)\right), \label{eq:wtheta-theory}
\end{eqnarray}
where the speed of light has been set to one, $\chi(z)$ is the comoving distance to a given redshift (in a flat universe, which is assumed throughout); $J_0$ is the Bessel function of order zero;  $H(z)$ is the Hubble expansion rate at redshift $z$; and $P_{\mathrm{NL}}(k;z)$ is the 3D matter power spectrum at redshift $z$ and wavenumber $k$ (which, in this Limber approximation, is set equal to $(l+1/2)/\chi$). Note that in Eq.~\ref{eq:wtheta-theory} we have assumed the bias to be constant within each bin, see Fig. 8 in \cite{MPPmethodology} for a test of this assumption. Again, all assumptions and approximations mentioned here have been shown to be inconsequential in \cite{MPPmethodology,simspaper}.
To model cross-correlation between redshift bins we simply change $n_{\rm g}^i(z)^2 \rightarrow n_{\rm g}^i(z) n_{\rm g}^j(z)$ and $(b^i)^2 \rightarrow b^i b^j$ in Eq.~\ref{eq:wtheta-theory}. 

Throughout this paper, we use the \textsc{CosmoSIS} framework \citep{cosmosis} to compute correlation functions, and to infer cosmological parameters. The evolution of linear density fluctuations is obtained using the \textsc{Camb} module \citep{camb} and then converted to a non-linear matter power spectrum  $P_{NL}(k)$ using the updated {\Halofit} recipe \citep{Takahashi2012}. 

The theory modeling we use assumes the Limber approximation, and it also neglects redshift space distortions. For the samples and redshift binning used in this paper, those effects start to become relevant at scales of $\theta \gtrsim 1 \deg$ \citep{2007MNRAS.378..852P,2010MNRAS.407..520N,2011MNRAS.414..329C}. In a companion paper \citep{MPPmethodology} it is explicitly shown that they have negligible impact in derived cosmological parameters given the statistical error bars of DES Y1. Concretely a theory data vector was produced with the exact (non-Limber) formula including redshift space distortions and was then analyzed with the baseline pipeline assumed here, and also in Y1COSMO. Figure 8 in \citep{MPPmethodology} shows that including or not including these contributions makes negligible impact in parameters such as $\Omega_m$ and $S_8$ for a LCDM Universe or $w$ in a $wCDM$ one. We also tested the impact of these effects on the fixed cosmology bias measurements in Section \ref{sec:galaxybias} and find them to be negligible. 
Hence in what follows, such terms are ignored for speed and simplicity. However future data analyses may need to account for these effects due to improved statistical uncertainty.

We model (and marginalize over) photometric redshift bias uncertainties as an additive shift $\Delta z^{i}$ in the \textsc{redMaGiC} redshift distribution $n_{\rm RM}^{i}(z)$ for each redshift bin $i$. 

\begin{equation}
n^{i}(z) = n^{i}_{\rm RM}(z - \Delta z^{i})
\end{equation}

The priors on the $\Delta z^{i}$ nuisance parameters, are measured directly using the angular cross correlation between the DES sample and a spectroscopic sample. These values are shown in Table \ref{bin_table2}, and the method is described in full in \cite{Cawthon17}. We use the same $\Delta z^{i}$ as Y1COSMO for all tests of robustness of the parameter constraints.

We also compare the measurements of $b^{i}$ to the same quantity measured by galaxy-galaxy lensing using the two-point correlation function $\gamma_{t}$ (see \cite{Prat17} for definition). We use the notation $b_{\times}^{i}$ for this measurement. The details of this measurement are described in \cite{Prat17} (hereafter Y1GGL). In order to take the off-diagonal elements of the covariance matrix between the two probes into account, we produce joint constraints from $w(\theta)$ and $\gamma_{t}$ at fixed cosmology (the mean of the Y1COSMO posterior), using different bias parameters for the two probes, and marginalizing over the same nuisance parameters as were used in the fiducial analysis of Y1COSMO (intrinsic alignments, source and lens photo$-z$ bias, and shear calibration). To test the consistency between the two probes we use the parameter $r$ which is,
\begin{equation}
r^{i} = \frac{ b^{i}_{\times} }{ b^{i} }. 
\end{equation} 
If $r \neq 1$, this indicates an inconsistency between the two bias measurements and would thus suggest a breakdown of our simple linear bias model. This test informs the choice of fixing $r=1$ in the Y1COSMO analysis. 

A combination of galaxy clustering and galaxy-galaxy lensing, provides a measurement of galaxy bias and $\sigma_{8}$ only if you assume that $r=1$. This test provides a measurement of $r$ which informs the choice of fixing $r=1$ in the Y1COSMO analysis. In principle, this test could also be performed by including the shear-shear correlation which also measures $\sigma_{8}$. 

\section{Data}
\label{sec:data}
\subsection{Y1 Gold}

\begin{figure*}
	\includegraphics[scale=0.6,trim={0 1.7cm 0 0}]{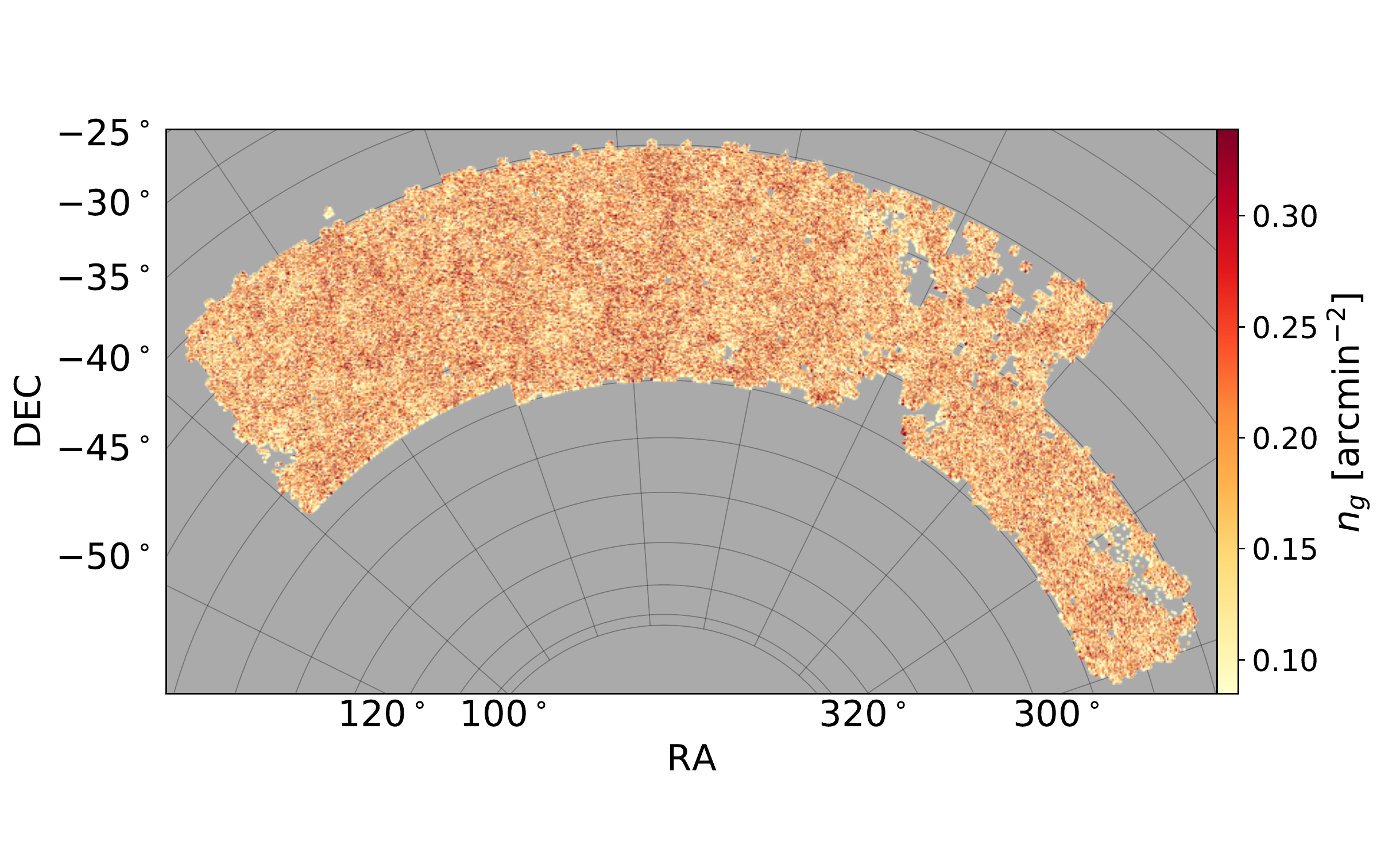}
	\caption{Galaxy distribution of the \textsc{redMaGiC} Y1 sample used in this analysis. The fluctuations represent the raw counts, without any of the corrections derived in this analysis. We have restricted the analysis to the contiguous region shown in the figure. The area is 1321 square degrees.}
    \label{galaxymap}
\end{figure*}

\begin{figure}
	\includegraphics[scale=0.55]{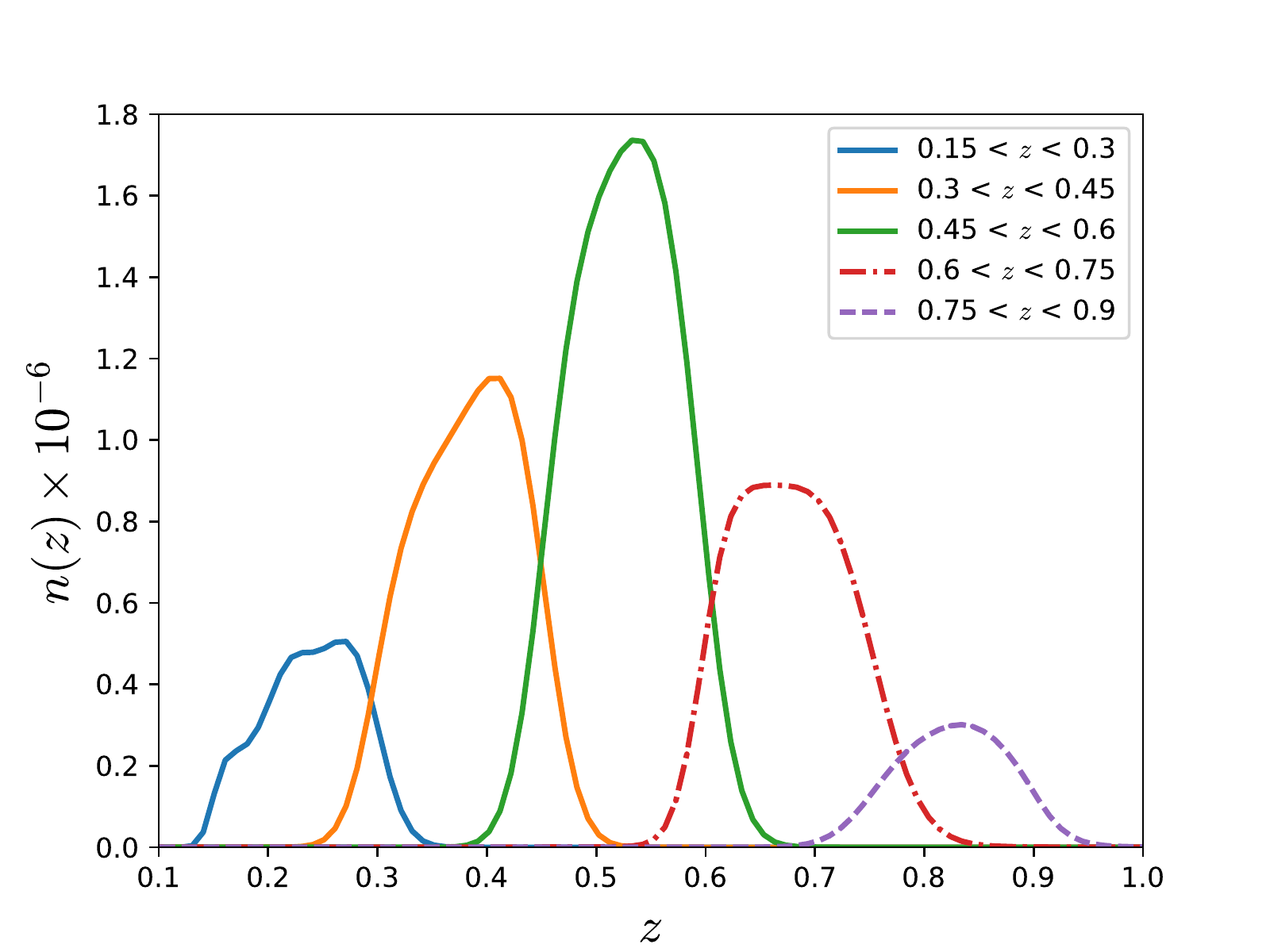}
	\caption{Redshift distribution of the combined \textsc{redMaGiC} sample in 5 redshift bins. They are calculated by stacking Gaussian PDFs with mean equal to the \textsc{redMaGiC} redshift prediction and standard deviation equal to the \textsc{redMaGiC} redshift error. Each curve is normalized so that the area of each curve matches the number of galaxies in its redshift bin. }
  \label{dndz}
\end{figure}

We use data taken in the first year (Y1) of DES observations \cite{Diehl14}. Photometry and `clean' galaxy samples were produced with this data as outlined by \cite{Y1gold} (hereafter denoted `Y1GOLD'). The outputs of this process represent the Y1 `Gold' catalog. Data were obtained over a total footprint of  $\sim$1800 deg$^2$; this footprint is defined by a {\sc Healpix} \citep{HealpixSoft} map at resolution $N_{\mathrm{side}}=4096$ (equivalent to 0.74 square arcmin) and includes only pixels with minimum exposure time of at least 90 seconds in each of the $g$,$r$,$i$, and $z$-bands, a valid calibration solution, as well as additional constraints (see Y1GOLD for details). A series of veto masks, including among others masks for bright stars and the Large Magellanic Cloud, reduce the area by $\sim 300$ deg$^2$, leaving $\sim$1500~deg$^2$ suitable for galaxy clustering study. We explain further cuts to the angular mask in Section \ref{sec:galsamples}. All data described in this and in subsequent sections are drawn from catalogues and maps generated as part of the DES Y1 Gold sample and are fully described in Y1GOLD.

\begin{figure*}
  \includegraphics[scale=0.4]{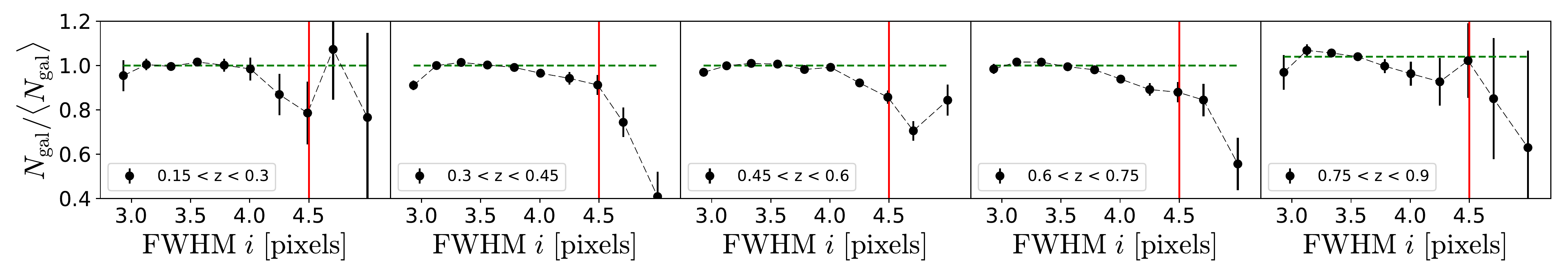}
  \includegraphics[scale=0.4]{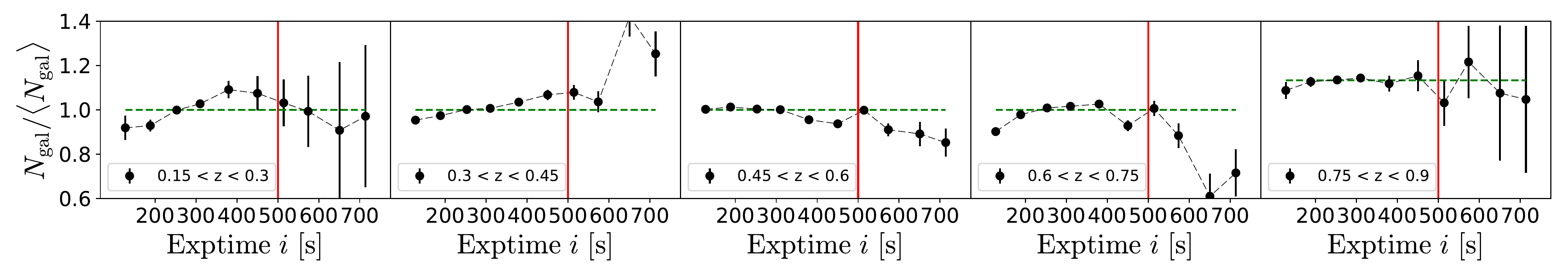}
  \caption{Correlations of volume-limited \textsc{redMaGiC} galaxy number density with seeing FWHM and exposure time before any survey property (SP; see text for more details) cuts (illustrated with the red vertical lines) were applied to the mask. In the absence of systematic correlations, the results obtained from these samples are expected to be consistent with no trend (the reference green dashed line). The cuts removed regions with $i$-band FWHM $> 4.5$ pixels and $i$-band exposure time $> 500s$ as these showed correlations that differed significantly from the mean ($>20\%$) or were not well fit by a monotonic function. No SP weights were used in this figure. }
  \label{1dsys}
\end{figure*}

\subsection{\textsc{redMaGiC} sample}

\label{sec:galsamples}

The galaxy sample we use in this work is generated by the \textsc{redMaGiC} algorithm, run on DES Y1 Gold data.  The \textsc{redMaGiC} algorithm selects Luminous Red Galaxies (LRGs) in such a way that photometric redshift uncertainties are minimized, as is described in~\cite{Rozo15}.  This method is able to achieve redshift uncertainties $\sigma_z/(1+z) < 0.02$ over the redshift range of interest.  The \textsc{redMaGiC} algorithm produces a redshift prediction $z_{\rm RM}$ and an uncertainty $\sigma_{z}$ which is assumed to be Gaussian. This sample was chosen instead of other DES photometric samples because of its small redshift uncertainty, which is obtained at the expense of number density. 

The \textsc{redMaGiC} algorithm makes use of an empirical red-sequence template generated by the training of the \textsc{redMaPPer} cluster finder~\citep{rykoff+14,rykoff+16}. As described in \citep{rykoff+16}, training of the red-sequence template requires overlapping spectroscopic redshifts, which in this work were obtained from SDSS in the Stripe 82 region~\cite{SDSSdr12} and the OzDES spectroscopic survey in the DES deep supernova fields~\cite{OzDESyear1}.

For the \textsc{redMaGiC} samples in this work, we make use of two separate versions of the red-sequence training.  The first is based on SExtractor \texttt{MAG\_AUTO} quantities from the Y1 coadd catalogs, as applied to \textsc{redMaPPer} in \citep{soergel+16}.  The second is based on a simultaneous multi-epoch, multi-band, and multi-object fit (\texttt{MOF})~(see Section 6.3 of Y1GOLD), as applied to \textsc{redMaPPer} \cite{Mcclintock17}. 
In general, due to the careful handling of the point-spread function (PSF) and matched multi-band photometry, the \texttt{MOF} photometry yields lower color scatter and, hence, smaller scatter in red-sequence photo-$z$s.  For each version of the catalog, photometric redshifts and uncertainties are primarily derived from the fit to the red-sequence template. In addition, an afterburner step is applied (as described in Section 3.4 of ~\cite{Rozo15}) to ensure that \textsc{redMaGiC} photo-$z$s and errors are consistent with those derived from the associated \textsc{redMaPPer} cluster catalog~\cite{Rozo15}.

\begin{table}
\begin{center}
\begin{tabular}{lccccc}
	\hline
	$z$ range & $L_{\rm min}/L_{*}$ & $n_{\rm gal}$ (arcmin$^{-2}$) & $N_{\rm gal}$ & Photometry\\
    \hline
	$0.15 < z < 0.3$ & $0.5 $ & 0.0134 & 63719 & MAGAUTO\\
	$0.3 < z < 0.45$ & $0.5 $ & 0.0344 & 163446 & MAGAUTO \\
	$0.45 < z < 0.6$ & $0.5 $ & 0.0511 & 240727 & MAGAUTO\\
	$0.6 < z < 0.75$ & $1.0 $ & 0.0303 & 143524 & MOF \\
	$0.75 < z < 0.9$ & $1.5 $ & 0.0089 & 42275 & MOF \\
  	\hline
\end{tabular}
\end{center}
\caption{Details of the sample in each redshift bin. $L_{\rm min}/L_{*}$ describes the minimum luminosity threshold of the sample, $n_{\rm gal}$ is the number of galaxies per square degree, and $N_{\rm gal}$ is the total number of galaxies. }
\label{bin_table1}
\end{table}

\begin{table}
\begin{center}
\begin{tabular}{lccc}
	\hline
	$z$ range \ & $b_{\rm fid}^{i}$ \ & $\Delta z^{i}$ \\
    \hline
	$0.15 < z < 0.3$ & 1.45 & Gauss ($0.008, 0.007$) \\
	$0.3 < z < 0.45$ & 1.55 & Gauss ($-0.005, 0.007$)\\
	$0.45 < z < 0.6$ & 1.65 & Gauss ($0.006, 0.006$)\\
	$0.6 < z < 0.75$ & 1.8  & Gauss ($0.00, 0.010$)\\
	$0.75 < z < 0.9$ & 2.0  & Gauss ($0.00, 0.010$)\\
  	\hline
\end{tabular}
\end{center}
\caption{Details of the fiducial parameters used for covariance and parameter constraints. Here, $b_{\rm fid}^{i}$ is the fiducial linear galaxy bias for bin $i$ applied to the Gaussian mock surveys we use to construct the covariance matrices. The $\Delta z^{i} \ {\rm prior}$ is a Gaussian prior applied to the additive redshift bias uncertainty. These were selected to match the analysis in (\citeauthor{DESY1cos}; Y1COSMO).
}
\label{bin_table2}
\end{table}

As described in \citep{Rozo15}, the \textsc{redMaGiC} algorithm computes color-cuts necessary to produce a luminosity-thresholded sample of constant co-moving density.  Both the luminosity threshold and desired density are independently configurable, but in practice higher luminosity thresholds require a lower density for good performance.  We note that in \citep{Rozo15} the co-moving density was computed with the central redshift of each galaxy ($z_{\rm RM}$).  For this work, the density was computed by sampling from a Gaussian distribution $z_{\rm RM}\pm\sigma_{z}$, which creates a more stable distribution near filter transitions.  This is the only substantial change to the \textsc{redMaGiC} algorithm since the publication of \citep{Rozo15}.

We use \textsc{redMaGiC} samples split into five redshift bins of width $\Delta z = 0.15$ from $z=0.15$ to $z=0.9$. We define our footprint such that the data in each redshift bin will be complete to its redshift limit across the entire footprint. To make this possible, we define samples based on a luminosity threshold. Reference luminosities are computed as a function of $L_*$, computed using a \citet{bruzualcharlot03} model for a single star-formation burst at $z=3$~\citep[See Section 3.2][]{rykoff+16}. Naturally, increasing the luminosity threshold provides a higher redshift sample as well as decreasing the comoving number density. Using a different luminosity threshold in each redshift bin allows us to maximize signal to noise while also providing a complete sample in each redshift bin. The details of these bins are given in Table~\ref{bin_table1}. 

The 5 redshift bins were chosen so that the width of the bins is significantly wider than the uncertainty on individual galaxy redshifts, but smaller than the difference between the maximum redshifts of the luminosity thresholds used.

In addition to the primary \textsc{redMaGiC} selection, we also apply a cut on the luminosity $L/{L_*} < 4$ as this was shown for DES Science Verification to reduce the stellar contamination in the sample, although this is mostly superfluous for Y1 Gold.  During testing, we find that the observational systematic relationships for the $0.5L_*$ sample, used for $z<0.6$, are minimized for the \texttt{MAG\_AUTO} sample, with a very minor impact on photo-$z$ performance.  For $L_* \geq 1.0$, used for $z>0.6$, we instead find that the observation systematic relationships are minimized for the \texttt{MOF} sample, and that the photo-$z$ performance is also improved.  Consequently, we use \texttt{MAG\_AUTO} for our $z<0.6$ sample and \texttt{MOF} for $z>0.6$. See Section~\ref{sec:systematics} for further discussion.

We build the area mask for the \textsc{redMaGiC} samples based on the depth information produced with the \textsc{redMaGiC} catalogs.  This information is provided by the $z_\mathrm{max}$ quantity, which describes the highest redshift at which a typical red galaxy of the adopted luminosity threshold (e.g. $0.5L_*$) can be detected at $10\sigma$ in the $z$-band, at $5\sigma$ in the $r$ and $i$-bands, and at $3\sigma$ in the $g$-band, as described in Section~3.4 of \citep{rykoff+16}. The quantity $z_\mathrm{max}$ varies from point to point in the survey due to observing conditions.  Consequently, we construct a $z_\mathrm{max}$ map, specified on a {\sc HealPix} map with $N_{\rm side} = 4096$.  In order to obtain a uniform expected number density of galaxies across the footprint, we only use regions for which $z_\mathrm{max}$ is higher than the upper edge of the redshift bin under consideration. The footprint is defined as the regions where this condition is met in all redshift bins. Thus, we only use pixels that satisfy each of the conditions where the $0.5L_{*}$ sample is complete to $z=0.6$, the $1.0L_{*}$ sample is complete to $z=0.75$, and the $1.5L_{*}$ sample is complete to $z=0.9$. We also restrict the analysis to the contiguous region shown in Figure \ref{galaxymap}.

An additional 1.6\% of the footprint is vetoed because it has extreme observing conditions. The selection of these cuts is detailed in Section \ref{sec:systematics}.

After masking and additional cuts, we obtain a total sample of 653,691 objects distributed over an area of 1321 square degrees, as shown in Fig. \ref{galaxymap}. The average redshift uncertainty of the sample is $\sigma_{z}/(1+z) = 0.0166$. The redshift distribution of each bin can be seen in Figure \ref{dndz}. The number of objects in each bin increases up to $z=0.6$ due to the increase in volume, and decreases at higher redshift due to the increased luminosity threshold.

\begin{figure*}
  \includegraphics[scale = 0.42]{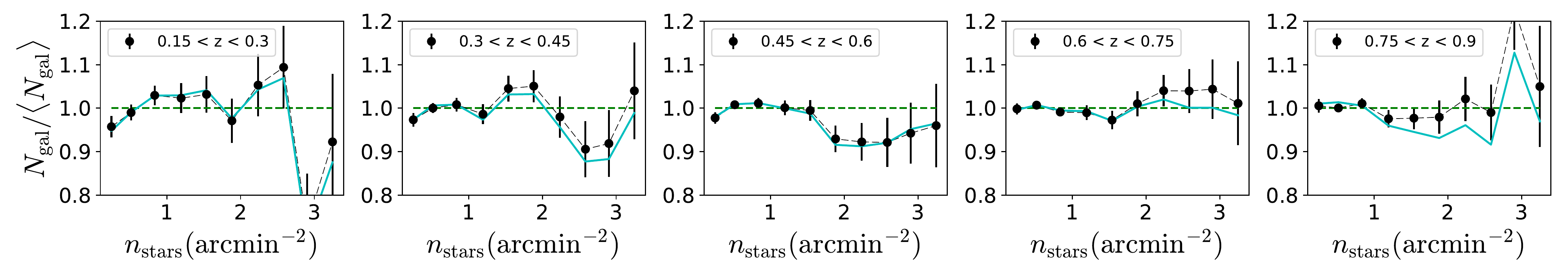}
  \caption{Galaxy number density divided by the mean number density across the footprint for each redshift bin, split by the number density of stars. The points with error-bars display the results for our 3$\Delta \chi^{2}(68)$ weighted sample, the cyan curves display the results without these weights. For the weighted sample, the $\chi^{2}$ of the line $N_{\rm gal}/\langle N_{\rm gal} \rangle = 1$ with the data points shown for each bin is 24.9, 16.0, 13.1, 6.6 and 10.9 with $N_{\rm d.o.f} = 10$. The $\Delta \chi^{2}$ between the null signal and a linear best fit is 0.99, 0.95, 0.24, 0.013, and 0.082. This does not meet either of the $\Delta \chi^{2}$ thresholds used in this analysis. We therefore see no evidence for stellar contamination or obscuration in this sample.}
  \label{fig:stars}
\end{figure*}

\begin{figure*}
  \includegraphics[scale = 0.5]{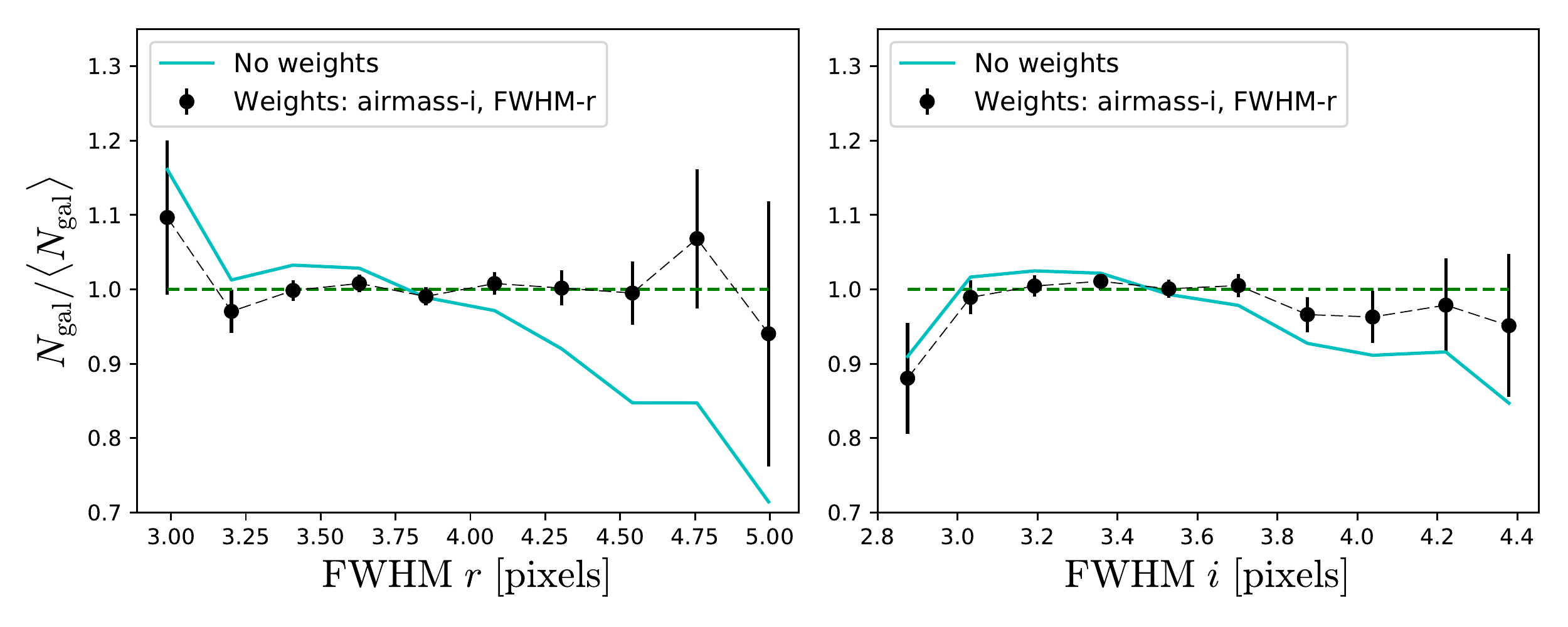}
  \caption{Galaxy number density in the highest redshift bin, $0.75 < z < 0.9$, as a function of two example SP maps, FWHM $r$-band and FWHM $i$-band. The black points correspond to the $3\Delta \chi^{2}(68)$ weighted sample, the cyan line is the unweighted sample. In this redshift bin, the SP maps used in the $3\Delta \chi^{2}(68)$ weights were Airmass $i$ and FWHM $r$. The left panel demonstrates the effect of the weights on the FWHM $r$ correlation. The right panel demonstrates that correlations with SP maps that were not included in the weights are still reduced due to correlations among the SP maps. The full set of SP correlations for the maps in Table \ref{weights_table} are shown in Appendix \ref{appendix:sys}. }
  \label{fig:examples}
\end{figure*}

\section {Analysis methods}
\label{sec:methods}

\subsection{Clustering estimators}
\label{sec:estimators}

We measure the correlation functions $w(\theta)$ using the Landy \& Szalay estimator \citep{LS}
\begin{equation}
\hat w(\theta) = \frac{DD - 2DR + RR}{RR} \, ,
\label{eqn:random_estimator}
\end{equation}
where $DD$, $RR$ and $DR$ are the number of pairs of galaxies from the galaxy sample $D$ and a random catalog $R$. This is calculated in 20 logarithmically separated bins in angle $\theta$ between 2.5 arcmin and 250 arcmin to match the analysis in Y1COSMO. We use 60 times more randoms than data. The pair-counting was done with the package {\tt tree-corr} \citep{2004MNRAS.352..338J} available at https://github.com/rmjarvis/TreeCorr.

We also calculate $w(\theta)$ on Gaussian random field realizations which are described in a pixelated map format. For these correlations we use a pixel-based estimator.
Using the notation of \cite{Crocce16}, the correlation between two maps $N_{1}$ and $N_{2}$ of mean values $\bar N_1, \bar N_2$ is estimated as
\begin{equation}
\!\!\!\!\!\! \hat w_{1,2}(\theta) = \sum^{N_{\rm pix}}_{i=1} \sum^{N_{\rm pix}}_{j=i} \frac{(N_{i, 1} - {\bar N_{1}})(N_{j,2} - {\bar N_{2}})}{{\bar N_{1}}{\bar N_{2}}} \Theta_{i,j} \, ,
\label{eqn:pixel_estimator}
\end{equation} 
where the sum runs through all pairs of the $N_{\rm pix}$ pixels in the footprint, $N_{i,1}$ is the value of the $N_{1}$ map in pixel $i$, and $\Theta_{i,j}$ is unity when the pixels $i$ and $j$ are separated by an angle $\theta$ within the bin size $\Delta \theta$. We have tested that the difference between the estimators in Equations \ref{eqn:random_estimator} and \ref{eqn:pixel_estimator} is negligible for this analysis. 

\subsection{Covariances}
\label{sec:covariances}

The fiducial covariance matrix we use for the $w(\theta)$ measurement is a theoretical halo model covariance, described and tested by \cite{MPPmethodology}. 
The covariance is generated using {\tt CosmoLike}~\cite{Krause:2016jvl}, and is computed by calculating the four-point correlation functions for galaxy clustering in the halo model. Additionally, an empirically determined correction for the survey geometry's effect on the shot-noise component has been added. The presence of boundaries and holes decrease the effective number of galaxy pairs as a function of pair separation, which in turn raises the error budget associated to shot noise over the standard uniform sky assumptions. This same covariance is used for the combined probes analysis and is detailed in Y1COSMO.

For the analysis of observational systematics and their correlation with the data, we use a set of 1000 mock surveys (hereafter `mocks') based on Gaussian random field realizations of the projected density field. These are then used to obtain an alternative covariance, which includes all the mask effects as in the real data. 
The mocks we use were produced using the following method. We first calculate, using {\sc \textsc{Camb}} \citep{camb}, the galaxy clustering power spectrum $C^{gg}_{i}(\ell)$, assuming the fiducial cosmology with fixed galaxy bias $b^{i}$ for each redshift bin $i$; the galaxy bias values are listed in Table \ref{bin_table2}. The angular power spectrum is then used to produce a full-sky Gaussian random field of $\delta_{g}$. We apply a mask to this field corresponding to the Y1 data, as shown in Fig.~\ref{galaxymap}. This is converted into a galaxy number count $N_{gal}$ as a function of sky position, with the same mean as the observed number count ${\bar N_{o}}$ in each redshift bin, using
\begin{equation}
N_{\rm gal} = {\bar N_{o}} (1 + \delta_{g}).
\end{equation}
Shot noise is finally added to this field by Poisson sampling the $N_{\rm gal}$ field. 

In order to avoid pixels with $\delta_{g} < -1$, which cannot be Poisson sampled, we follow the method used by \cite{MPPmethodology}: before Poisson sampling, we multiply the density field by a factor $\alpha$, where $\alpha < 1$; we then rescale the number density $n_{\rm gal}$ by $1/\alpha^{2}$ in order to preserve the ratio of shot-noise to sample variance; we then rescale the measured $w(\theta)$ by $1/\alpha^{2}$ to obtain the unbiased $w(\theta)$ for each mock. This procedure is summarized by
\begin{eqnarray}
\delta_{g} &\rightarrow& \alpha \delta_{g} \, ,  \\
n_{\rm gal} &\rightarrow& n_{\rm gal}/\alpha^2 \, , \\
w(\theta) &\rightarrow& w(\theta)/\alpha^{2} \, .
\end{eqnarray}

We then use these mocks to estimate statistical errors in galaxy number density as a function of potential systematics. Alternatively we ``contaminate'' each of the 1000 mocks with survey properties as discussed in Section \ref{sec:systematics} to assess the impact of systematics on the $w(\theta)$ covariance. Note that these mocks would not be fully realistic for $w(\theta)$ covariance and cosmological inference as they are basically Gaussian realizations. These mocks allow us to quantify significances (i.e., a $\Delta\chi^{2}$) to null tests, which are a necessary step in our analysis. Further, given such a large number of realizations we are able to obtain estimates of both the impact of the systematic correction on the resulting statistical uncertainty and any bias imparted by our methodology to well below 1$\sigma$ significance (e.g., given 1000 mocks, a 0.1$\sigma$ bias can be detected at 3$\sigma$ significance).

\section{Systematics}
\label{sec:systematics}

\subsection {Survey property (SP) maps}
\label{sp_maps}

The number density of galaxies selected based on their imaging is likely to fluctuate with the imaging quality due to fluctuations in the noise (e.g., Malmquist bias) and limitations in the reduction pipeline. Such fluctuations can imprint the structure of certain survey properties onto the galaxy field, thereby producing a non-cosmological signal. In order to quantify the extent of these correlations and remove their effect from the two-point function, maps of DES imaging properties were produced using the methods described in Ref.~\cite{LeistedtMAPS}. We consider the possibility that depth, seeing, exposure time, sky brightness and airmass, in each band $griz$, affect the density of galaxies we select. 

In total, we consider 21 survey property maps. We refer to these as SP maps from here on: 
\begin{itemize}
\item depth: the magnitude limit at which we expect to be able to detect a galaxy to 10$\sigma$ significance;
\item seeing FWHM: the full width at half maximum of the PSF of a point source;
\item exposure time: total exposure time in a given band;
\item sky brightness: the brightness of the sky, e.g., due to background light or the Moon phase;
\item airmass: the amount of atmosphere a source has passed through, normalized to be 1 when pointing at zenith.
\end{itemize}
\noindent Where relevant, we use the weighted average quantity over all exposures contributing to a given area. 

We also consider Galactic extinction and stellar contamination (or obscuration \cite{Ross11}) as potential systematics. The stellar density map was created by selecting moderately bright, high confidence stars. Using the notation of Y1GOLD, this selection is \texttt{MODEST\_CLASS} $=2$ with $18.0<i<20.5$, \texttt{FLAGS\_GOLD} $= 0$, and \texttt{BADMASK} $\le 2$. We also include an additional color cut of $0.0 < g-i < 3.5$ and $g-r>0$. These stars were binned in pixels with $N_{\rm side}=512$ (equivalent to 47 square arcmin), and the corresponding area for each pixel was computed at higher resolution ($N_{\rm side}=4096$) from the Y1 Gold footprint and pixel coverage fraction, as well as the bad region mask. Together, this yields the number of moderately bright stars per square degree that can be used to cross-correlate galaxies with stellar density. Using \texttt{MODEST\_CLASS} to select stars means this map could potentially contain DES galaxies. For this reason, we test for correlations with the astrophysical maps separately to the SP maps. As we find no correlation between stellar density and galaxy density, we do not take this contamination into account. For Galactic extinction, we use the standard map from \cite{SFD}.

\subsection {Systematic corrections}
\label{sys_correction}

This section describes the method used to identify and correct for observational systematics. We also discuss the uncertainty on this correction and its impact on the $w(\theta)$ covariance. Our approach is to first identify maps that are correlated with fluctuations in the galaxy density field at a given significance.
We then correct for the contamination using weights to be applied to the galaxy catalog. 

As demonstrated by \cite{Elsner16}, when testing a large number of maps one expects there to be some amount of covariance between the maps and the true galaxy density field due to chance. Consequently, it is possible to over-correct the galaxy density field using the type of methods employed in this work.  To limit this effect, we do not correct for all possible maps, and limit ourselves to those maps that are detected to be correlated with the galaxy density field at high significance (above a given threshold). We test the robustness of the results on our choice of threshold in Section \ref{sec:threshold} and we test for biases due to over-correction in Section \ref{falsebias}. The end result of our procedure is a measurement of $w(\theta)$ that is free of systematics above a given significance (in our concrete case a galaxy density free of two sigma correlations with SP maps, as defined below, and visualized in Fig.~\ref{weights_table}) and that can be directly utilized in combination with weak lensing measurements for cosmological analyses.

We identify the most significant SP maps as follows.  First, given an SP map of some quantity $s$, we identify all pixels in some bin $s \in [s_{\min}, s_{\max}]$.  We then compute the average density of galaxies in these pixels.  By scanning across the whole range of possible $s$-values for the SP map, we can directly observe how the galaxy density field scales with $s$. Examples of these type of analyses are shown in Figs. \ref{1dsys}, \ref{fig:stars} and \ref{fig:examples}.

We first remove regions of the footprint that display either especially significant ($>20$\%) changes in galaxy density from the mean, or are poorly fit by a monotonic function. These regions are defined from the cuts shown in Fig \ref{1dsys}. We remove regions of the footprint with $i$-band FWHM $> 4.5$ and $i$-band exposure time $> 500s$. These cuts remove 1.6\% of the Y1 area. 

\begin{table}
\begin {center}
\begin{tabular}{lllll}
\hline
	$z$ range 	& Maps included in 				& Maps included in \\
	 			& 3$\Delta\chi^2(68)$ weights		& 2$\Delta\chi^2(68)$ weights \\ \hline
	$0.15 < z < 0.3$ 	& Depth ($r$) 			& Exptime ($i$)\\
	 				 	& 			 			& FHWM ($z$)\\
     					& 				 		& FWHM ($r$)\\
     					& 				 		& Airmass ($z$)\\ \hline
	$0.3 < z < 0.45$ 	& Depth ($g$) 			& Depth ($g$)\\ \hline
	$0.45 < z < 0.6$ 	& FWHM ($z$) 			& FWHM ($z$)\\
    					& Exptime ($g$) 		& Exptime ($g$)\\
    					& FWHM ($r$) 			& FWHM ($r$)\\
                        & Skybright ($z$) 		& Skybright ($z$)\\
                        & 				 		& Depth ($i$)\\ \hline
	$0.6 < z < 0.75$ 	& FWHM ($gri$) PCA-0 	& FWHM ($gri$) PCA-0\\
    					& Skybright ($r$) 		& Skybright ($r$)\\
    					& FWHM ($z$) 			& FWHM ($z$)\\ 
                        & 			 			& Exptime ($i$)\\
                        & 			 			& Exptime ($z$)\\ \hline
	$0.75 < z < 0.9$ 	& Airmass ($i$) 		& Airmass ($i$)\\
                        & FWHM ($r$) 			& FWHM ($r$)\\
                        & 			 			& FWHM ($g$)\\
                        \hline
\end{tabular}
\end{center}
\caption{List of the maps used in the SP weights. Each of these has been determined to impart fluctuations in our galaxy sample at $>3\Delta\chi^2(68)$ or $>2\Delta\chi^2(68)$ significance. The weights were applied serially for each map in the order shown, starting from the top of the table. `FWHM' refers to the full-width-half-maximum size of the PSF. The photometric band of each SP map is in parentheses.}
\label{weights_table}
\end{table}

\begin{figure*}
  \centering
  \includegraphics[scale=0.5]{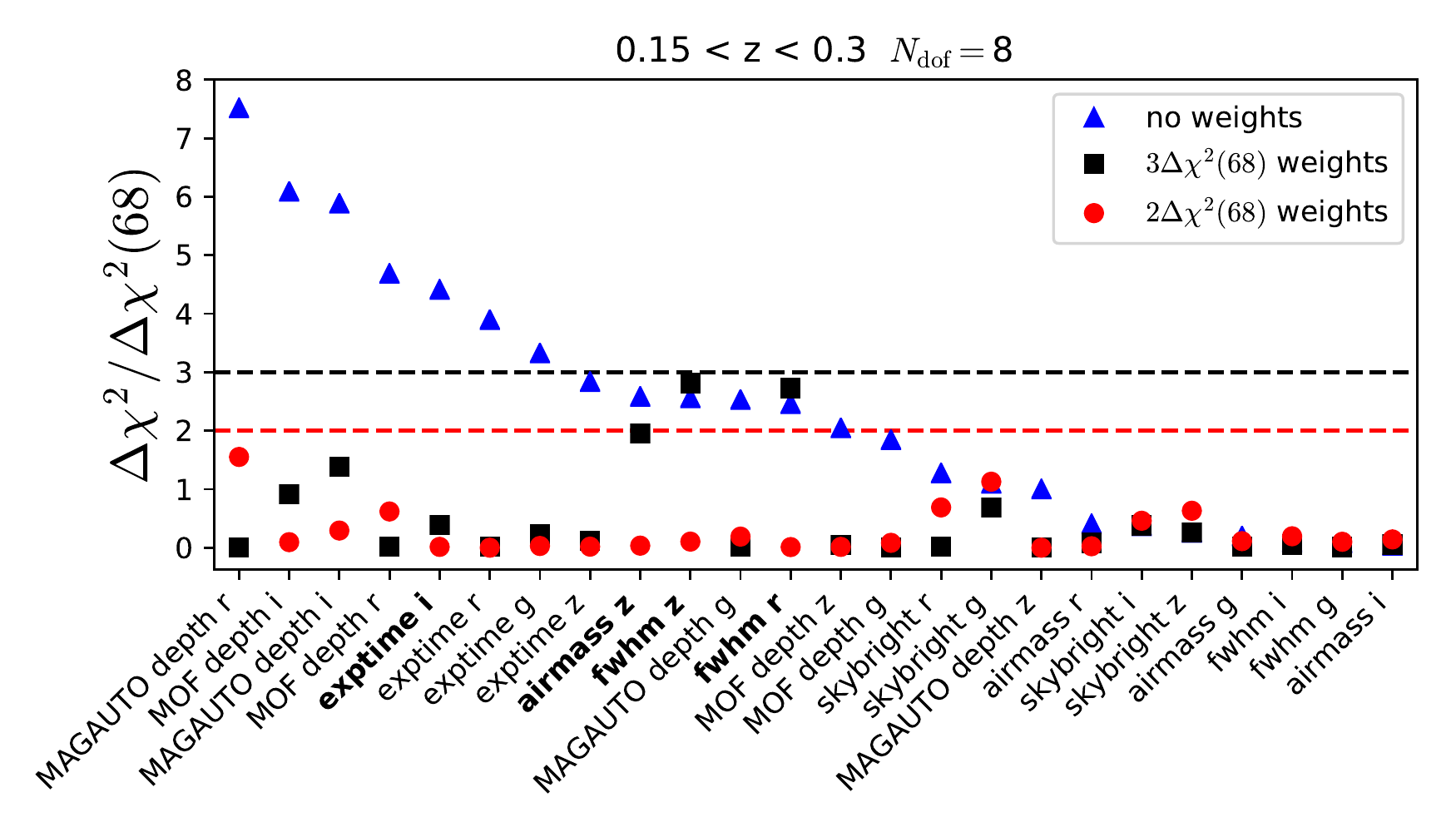}
  \includegraphics[scale=0.5]{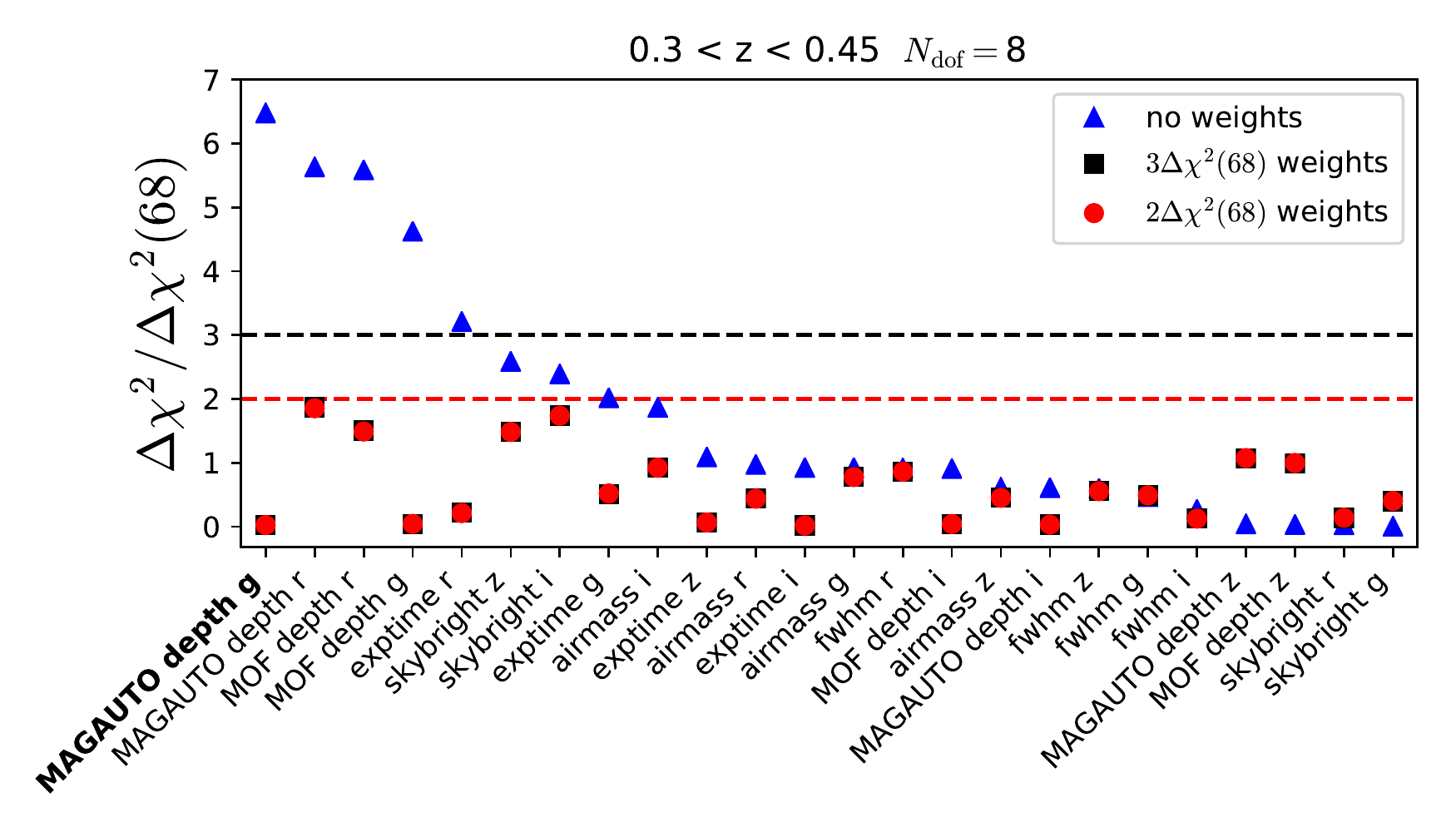}
  \hspace{0mm}
  \includegraphics[scale=0.5]{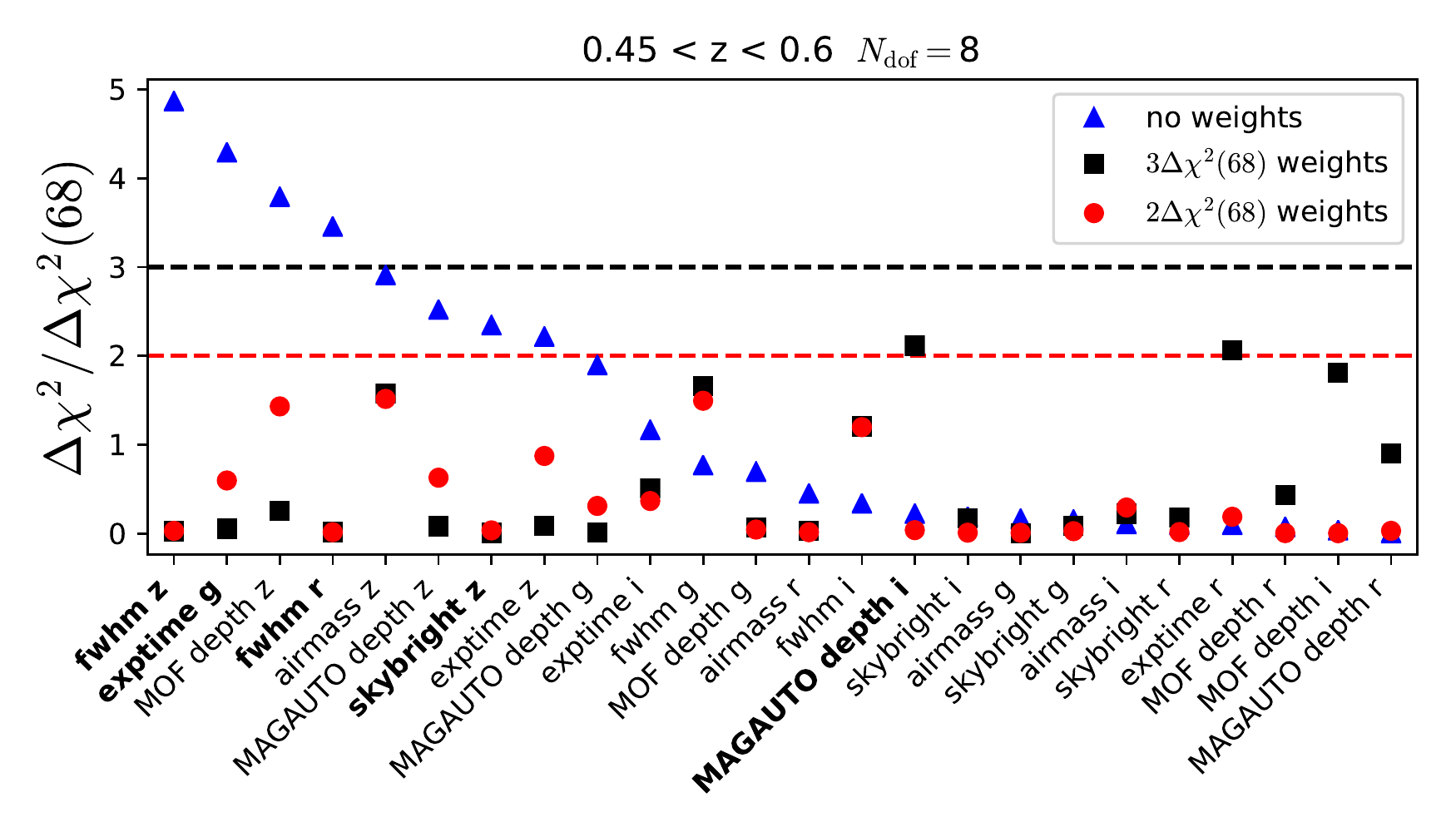}
  \includegraphics[scale=0.5]{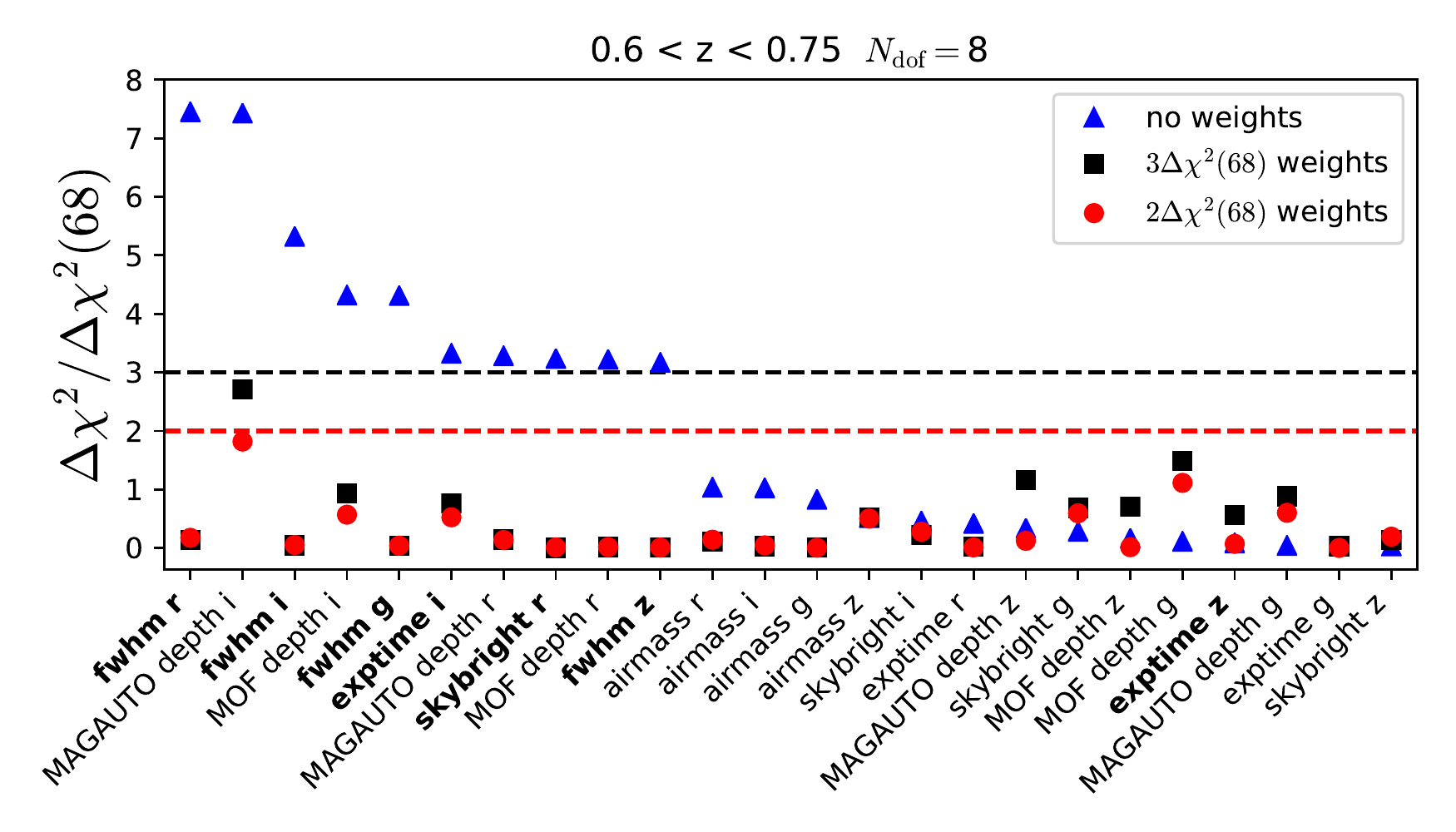}
  \hspace{0mm}
  \includegraphics[scale=0.5]{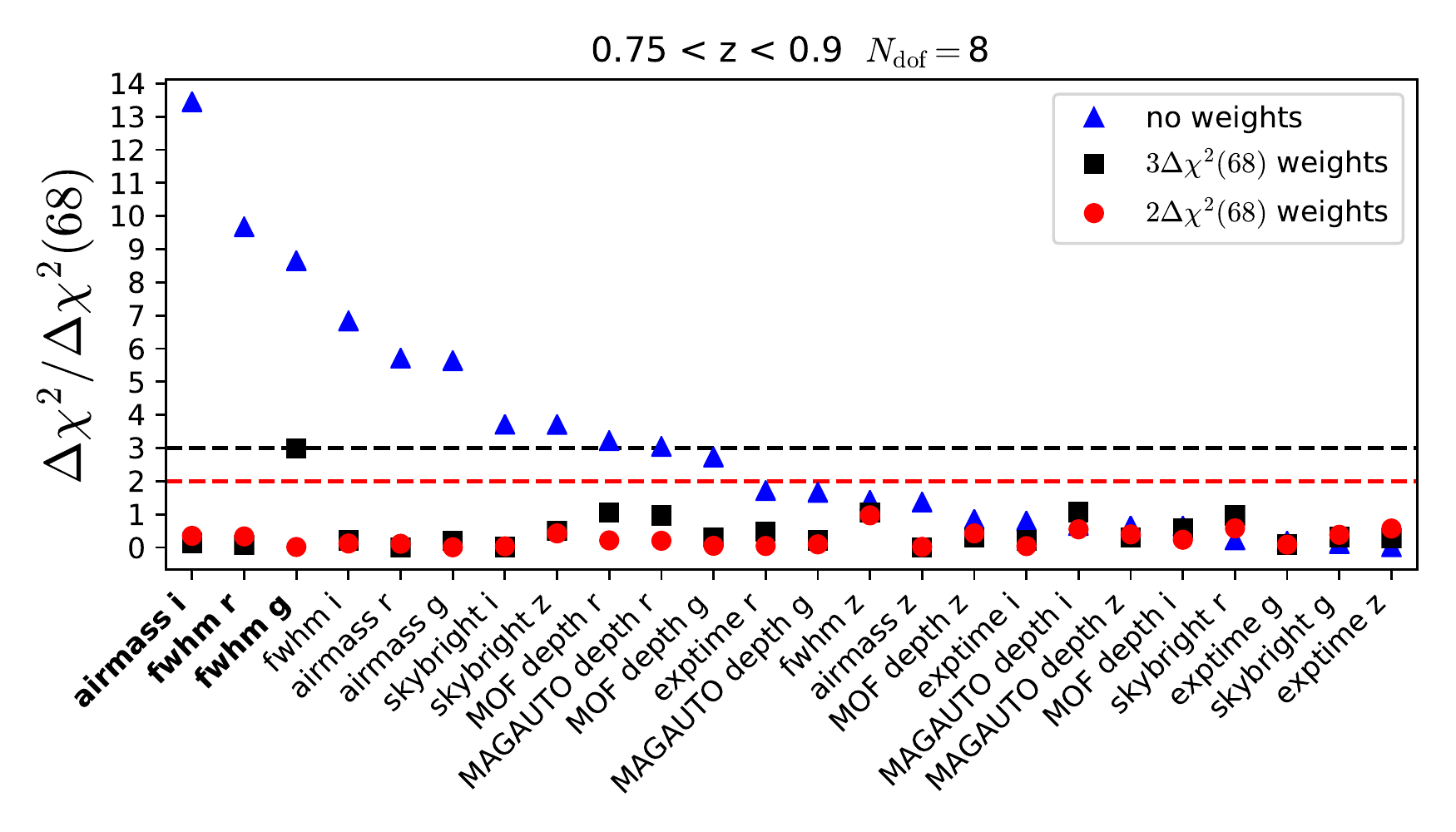}
  \caption{The significance of each systematic correlation. The significance is calculated by comparing the $\Delta \chi^{2}$ measured on the data to the distribution in the mock realizations. We find the 68th percentile $\Delta\chi^2$ value, labeling it $\Delta\chi^2(68)$, for each map obtained from the mock realizations. We quote the significance for the relationship obtained on the data as $\Delta\chi^2/\Delta\chi^2(68)$. Weights are applied for the SP map with the largest significance, with the caveat that we do not correct for both depth and the components of depth (e.g. exposure time, PSF FWHM) in the same band. For example, in the bin $0.15 < z < 0.3$, correcting for $r$-band depth (the most significant contaminant) did not remove all the $r$-band correlations with $\Delta\chi^2/\Delta\chi^2(68) >2$, so is not included in the final $2\Delta\chi^2/\Delta\chi^2(68)$ weights. 
This is repeated iteratively until all maps are below a threshold significance, shown here for thresholds of $2\Delta\chi^2/\Delta\chi^2(68)$ and $3\Delta\chi^2/\Delta\chi^2(68)$. The $x$ axis is shown in order of decreasing significance for the unweighted sample. The labels in {\bf bold} are the SP maps included in the $2\Delta\chi^2/\Delta\chi^2(68)$ weights. In the second redshift bin, $0.3 < z < 0.45$, the $3\Delta\chi^2/\Delta\chi^2(68)$ and $2\Delta\chi^2/\Delta\chi^2(68)$ weights are the same because correcting for only $g$-band depth removes all correlations with $\Delta\chi^2/\Delta\chi^2(68)>2$.  }
  \label{null_test_plot}
\end{figure*}

After cutting the footprint, we determine which SP maps most significantly correlate with the data by fitting a linear function to each number density relationship. We minimize a $\chi^2_{\mathrm{model}}$ where the model is $N_{\mathrm{gal}} \propto A \, s + B$. We determine the significance of a correlation based on the difference in $\chi^{2}$ between the best-fit linear parameters, and a null test of $N_{\rm gal}/ \langle N_{\rm gal} \rangle = 1$,
\begin{equation}
\Delta \chi^{2}=\chi^{2}_{\rm null}-\chi^{2}_{\rm model} \, .
\end{equation}

The $\Delta \chi^{2}$ is then compared to the same quantity measured on the Gaussian random fields described in Section \ref{sec:covariances}. We then label each potential systematic to be significant at $1 \sigma$ if the $\Delta \chi^{2}$ measured on the data is larger than 68\% of the mocks respectively. We denote this threshold as $\Delta \chi^{2}(68)$ and quote significances as $\Delta\chi^{2}/\Delta \chi^{2}(68)$; the square-root of this number should roughly correspond to the significance in terms of $\sigma$. 
Some examples of these tests for the observational systematics can be seen in Fig.~\ref{null_test_plot}. The full set of tests can be seen in Appendix \ref{appendix:sys}. We see no significant correlation with stellar density in the sample, as shown in Fig.~\ref{fig:stars}. Similarly, we find no correlations with Galactic extinction. Thus, our main tests are against SP maps, which are particular to DES observations.

\begin{figure*}
  \includegraphics[scale= 0.6]{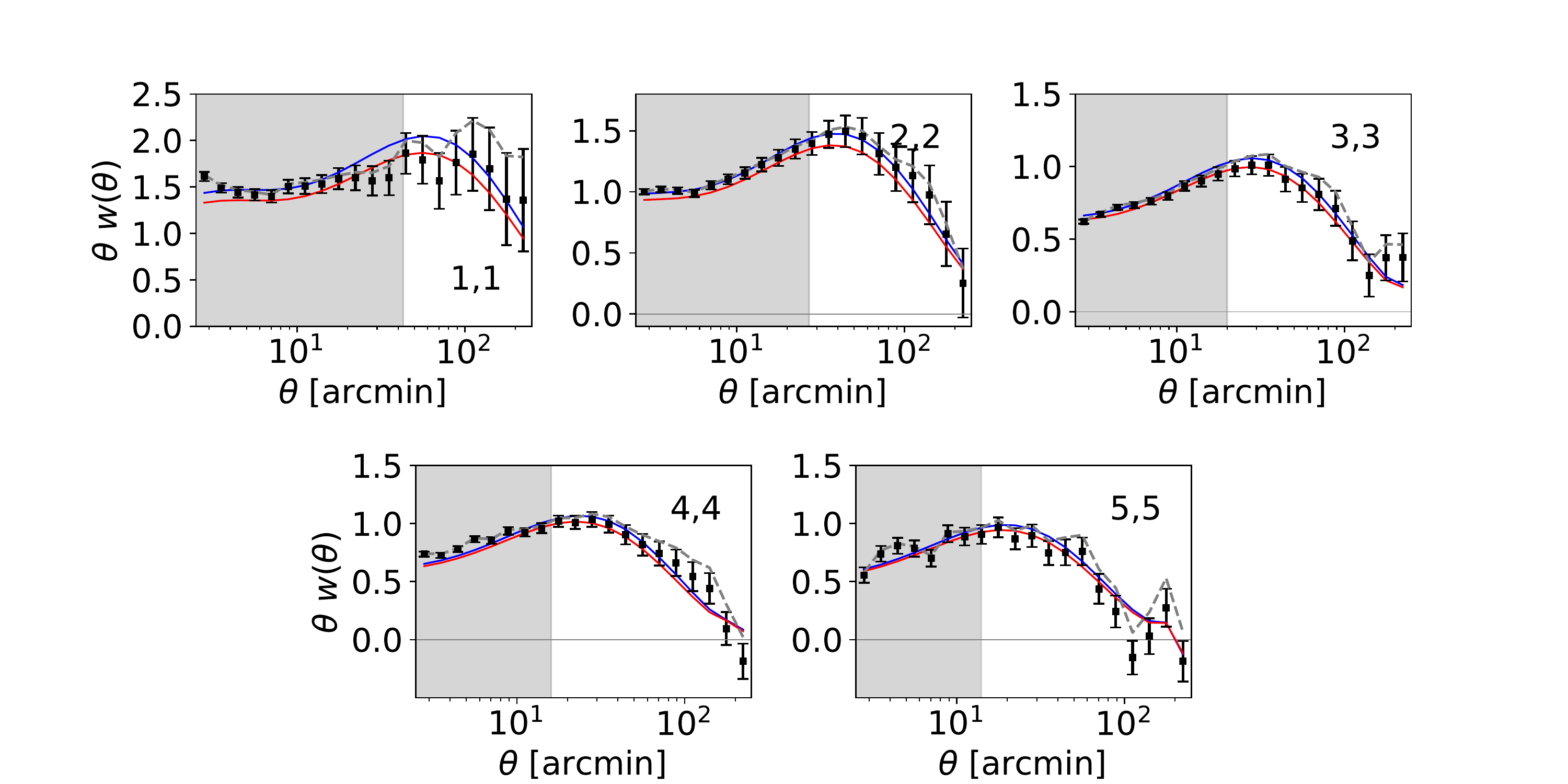}
	\caption{Two-point correlation functions for the fiducial analysis in each of the 5 redshift bins. These panels show the auto-correlation used in Y1COSMO and the galaxy bias measurements presented in this work.  A correction for correlations with survey properties is applied according to the methodology in Section \ref{sec:systematics}. The grey dashed line is the correlation function calculated without the SP weights. The black points use the $2\Delta\chi^2(68)$ weights. We show correlations down to $\theta = 2.5'$ to highlight the goodness of the fit towards small scales, but data points within grey shaded regions have not been used in bias constraints or the galaxy clustering part of Y1COSMO. That scale cut has been set in co-moving coordinates at $8\,{\rm Mpc}\,h^{-1}$. The solid red curve is the best-fit model using only the $w(\theta)$ auto-correlations at fixed cosmology, using $\Delta z^{i}$ priors from \cite{Cawthon17}. The solid blue curve is the best-fit model from the full cosmological analysis in Y1COSMO. }
  \label{wth}
\end{figure*}

\begin{table}
\begin{center}
\begin{tabular}{p{2.5cm} m{2.5cm} m{2.5cm} }
	\hline
	$z$ range & $b^{i}(\sigma_{8}/0.81)$ & $r^{i}$\\
    \hline
	$0.15 < z < 0.3$ & $ 1.40 \pm 0.072$ & $1.10 \pm 0.08$ \\
	$0.3 < z < 0.45$ & $ 1.60 \pm 0.051$ & $0.97 \pm 0.06$ \\
	$0.45 < z < 0.6$ & $ 1.60 \pm 0.039$ & $0.91 \pm 0.08$ \\
	$0.6 < z < 0.75$ & $ 1.93 \pm 0.045$ & $1.02 \pm 0.13$ \\
	$0.75 < z < 0.9$ & $ 1.98 \pm 0.070$ & $0.85 \pm 0.28$ \\
  	\hline
\end{tabular}
\end{center}
\caption{The measurements of galaxy bias $b^{i}$ and the ratio of bias from clustering and galaxy-galaxy lensing $r^{i}$ for each redshift bin $i$, calculated with cosmological parameters fixed at the mean of the Y1COSMO posterior, varying only bias and nuisance parameters with lens photo$-z$ priors from \cite{Cawthon17}.}
\label{bias_table}
\end{table}

Once we identify the most significant contaminant SP maps, we define weights to be applied to the galaxy sample in order to remove the dependency, following a method close to that of the latest LSS survey analysis \cite{Ross12,Ross17,Kwan17DES, Blake2010}. 
Note however that we identify systematics using a rigorous $\chi^2$ threshold significance criteria, based on a large set of Gaussian realizations, which to our knowledge was not done before.

For this method we apply the following steps to each redshift bin separately. The correlation with a systematic $s$ is fitted with a function $N_{\rm gal}/\langle N_{\rm gal} \rangle = F_{\rm sys}(s)$. 

For depth and airmass, the function used was a linear fit in $s$. For exposure time and sky brightness, the function was linear in $\sqrt s$, as this is how these quantities enter the depth map. For the seeing correlations, we fit the model
\begin{eqnarray}
\!\!\!\! N_{\rm gal}/\langle N_{\rm gal} \rangle  &=& F_{\rm sys} \, (s_{\rm FWHM})\nonumber   \\ 
\!\!\!\! F_{\rm sys} \, (s_{\rm FWHM}) & = & A \left[ 1 - {\rm erf} \left( \frac{s_{\rm FWHM} - B}{\sigma} \right) \right] \, ,
\end{eqnarray}
where $s_{\rm FWHM}$ is the seeing full-width half-max value, and  $A$, $B$ and $\sigma$ are parameters to be fitted. This functional form matches that applied to BOSS \citep{Anderson14,Ross17}; we believe it is thus the expected form when morphological cuts are applied to reject stars (as this is what causes the relationship for BOSS).

Each galaxy $i$ in the sample is then assigned a weight $1/F_{\rm sys}(s_{i})$ where $s_{i}$ is the value of the systematic at the galaxy's location on the sky. This weight is then used when calculating $w(\theta)$ and in all further null tests. 

In this sample we find evidence of multiple systematics at a significance of $\Delta\chi^{2}/\Delta \chi^{2}(68) > 3$, some of which are correlated with each other. To account for this, we first apply weights for the systematic with the highest $\Delta\chi^{2}/\Delta \chi^{2}(68)$. Then, using the weighted sample, we remeasure the significance of each remaining potential systematic and repeat the process until there are no systematics with a significance greater than a $\Delta\chi^{2}/\Delta \chi^{2}(68) = 3$ threshold. The final weights are the product of the weights from each required systematic. We also produce weights using a threshold of $\Delta\chi^{2}/\Delta \chi^{2}(68) = 2$, allowing us to determine if using a greater threshold has any impact on our clustering measurements. We refer to these weights as the $3\Delta \chi^{2}(68)$ and $2\Delta \chi^{2}(68)$ weights respectively. 
 
The final weights used in this sample are described in Table \ref{weights_table}. The SP maps are either the depth or properties that contribute to the depth (e.g. holding everything else fixed, a longer exposure time will result in an increased depth). Thus, in bins where multiple SP weights were required, we avoided correcting for both depth and SPs that contribute to the depth in the same band. In these cases, we weight for only the SPs that contribute to the depth. Fig.~\ref{1dsys_story} shows the correlation between the sample density and the SP maps used in Table \ref{weights_table}, both with and without weights. 

Fig.~\ref{null_test_plot} summarizes the results of our search for contaminating SPs, for each redshift bin. The blue points show the significance for each map, prior to the application of any weights. The black and red points display the significance after applying the $3\Delta \chi^{2}(68)$ and $2\Delta \chi^{2}(68)$ weights respectively. 
In Section \ref{sec:robust}, we will test our results with both choice of weights and whether to expect any bias from over-correction from either choice.

\begin{figure*}
  \includegraphics[width=\textwidth]{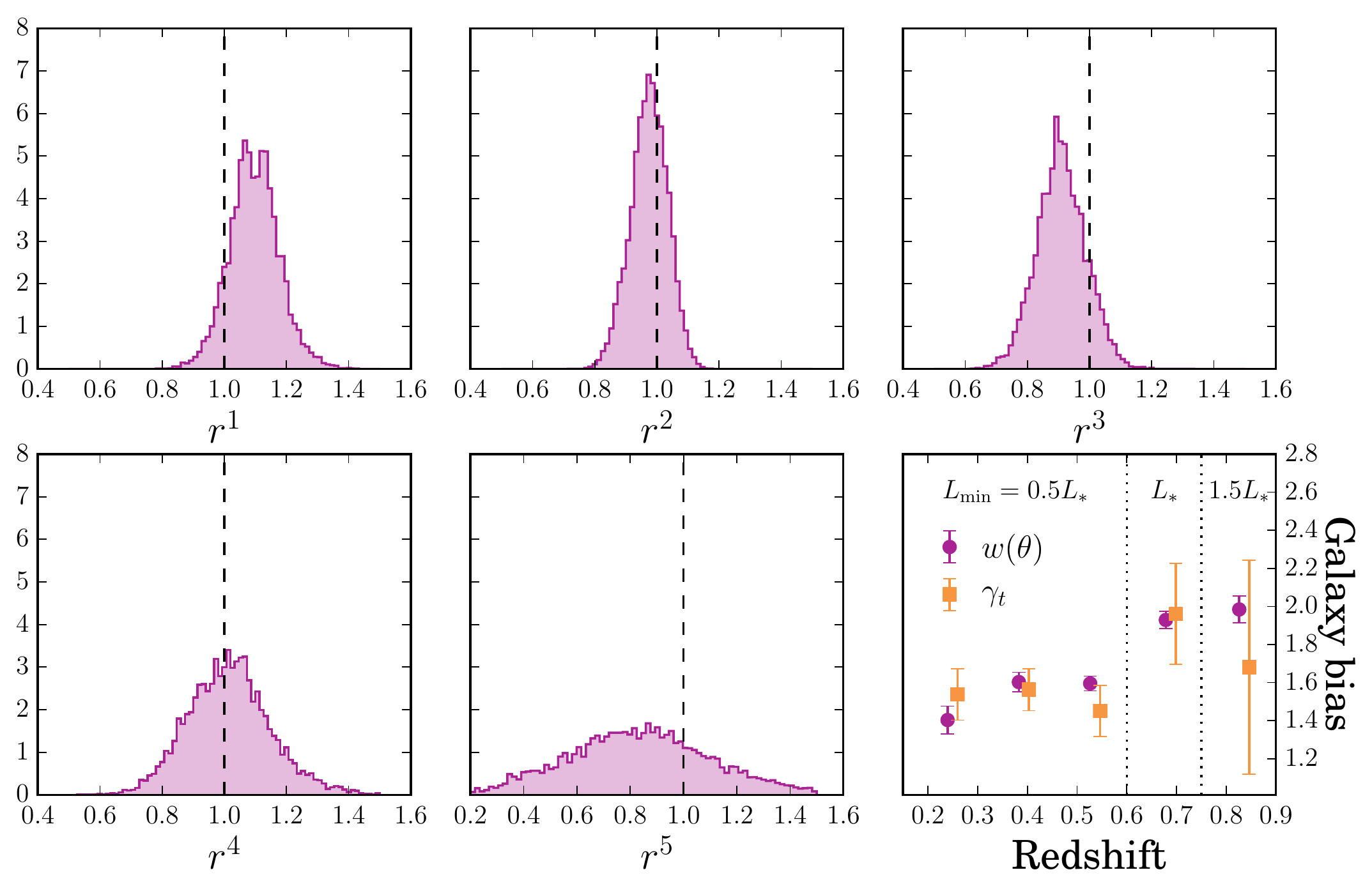}
  \caption{Constraints on the ratio, $r$, of galaxy bias measured on $w(\theta)$  and measured from the galaxy-galaxy lensing signal (see \cite{Prat17} denoted Y1GGL in the text) in each redshift bin. The histograms show the posterior distributions of $r^i$ from an MCMC fit for each $z$ in $i$. The bottom-right panel displays the individual measurements for each bin (purple for our $w(\theta)$ measurements and orange for those obtained in Y1GGL). All cosmological parameters were fixed at the DES Y1COSMO posterior mean values, and all nuisance parameters were varied as in Y1COSMO. The constraints were calculated using the full Y1COSMO covariance matrix, so the covariance between the two probes has been taken into account.  We see no significant evidence for $r \neq 1$ within the errors.}
  \label{r_plot}
\end{figure*}

When $F_{\rm sys}(s)$ is a linear function, the method described above, hereby referred to as the weights method, should be equivalent to the method used in \cite{Ho12,Crocce16}. This has been shown in \cite{Kwan17DES} for the DES science verification redMaGiC sample. 

The impact of the SP weights on the $w(\theta)$ measurement can be seen in Figure \ref{wth}. The dashed line displays the measurement with no weights applied. One can see that in all redshift bins, the application of the SP weights reduces the clustering amplitude and that the effect is greatest on large scales. This is consistent with expectations (see, e.g. Ref.~\cite{Ross11}).

\section{Results: Galaxy Bias and Stochasticity}
\label {sec:galaxybias}

In this section we present measurements of galaxy bias $b^{i}$ and stochastic bias $r^{i}$. The amplitude of the galaxy clustering signal is determined by the combination of parameters $(b^{i}\sigma_{8})^{2}$. Equivalently the galaxy-galaxy lensing signal $\gamma_{t}$ is sensitive to $b_{\times}^{i}(\sigma_{8})^{2}$. In the Y1COSMO combined probes analysis, cosmic shear provides a measurement of $\sigma_{8}$ meaning that galaxy clustering and galaxy galaxy lensing can each provide an independant measurment of galaxy bias (and therefore one could measure $r$). In this analysis we fix $\sigma_{8}$ at the mean of the Y1COSMO postierior ($\sigma_{8} = 0.81$) to measure $b^{i}(\sigma_{8}/0.81)$ from clustering and $r^{i}$ from $\gamma_{t}$. This provides a cosmology dependent measurement of bias from clustering alone, and test of the assumption $r=1$ in Y1COSMO.    

The $w(\theta)$ auto-correlation functions of the \textsc{redMaGiC} galaxy sample are shown in Figure \ref{wth}. We show the auto-correlation calculated with and without a correction for observational systematics, as described in Sec \ref{sec:systematics}. A minimum angular scale $\theta^{i}_{\rm min}$ has been applied to each redshift bin $i$. These were chosen to be $\theta^{1}_{\rm min}=43'$, $\theta^{2}_{\rm min}=27'$, $\theta^{3}_{\rm min}=20'$, $\theta^{4}_{\rm min}=16'$, and $\theta^{5}_{\rm min}=14'$ to match the analysis in Y1COSMO. These minimum angular scales, varying with redshift, correspond to a single minimum co-moving scale $R = 8\,{\rm Mpc} h^{-1}$ such that $\theta^i_{min} = R / \chi(\langle z^i \rangle)$, where $\langle z^i \rangle$ is the mean redshift of galaxies in bin $i$ \cite{MPPmethodology}. The scale was chosen so that a significant non-linear galaxy bias or baryonic feedback component to the Y1COSMO data vector would not bias the cosmological parameter constraints. 

\begin{figure*}
  \includegraphics[trim= 0.0cm 0cm 0.0cm 0cm, clip, totalheight=0.3\textwidth]{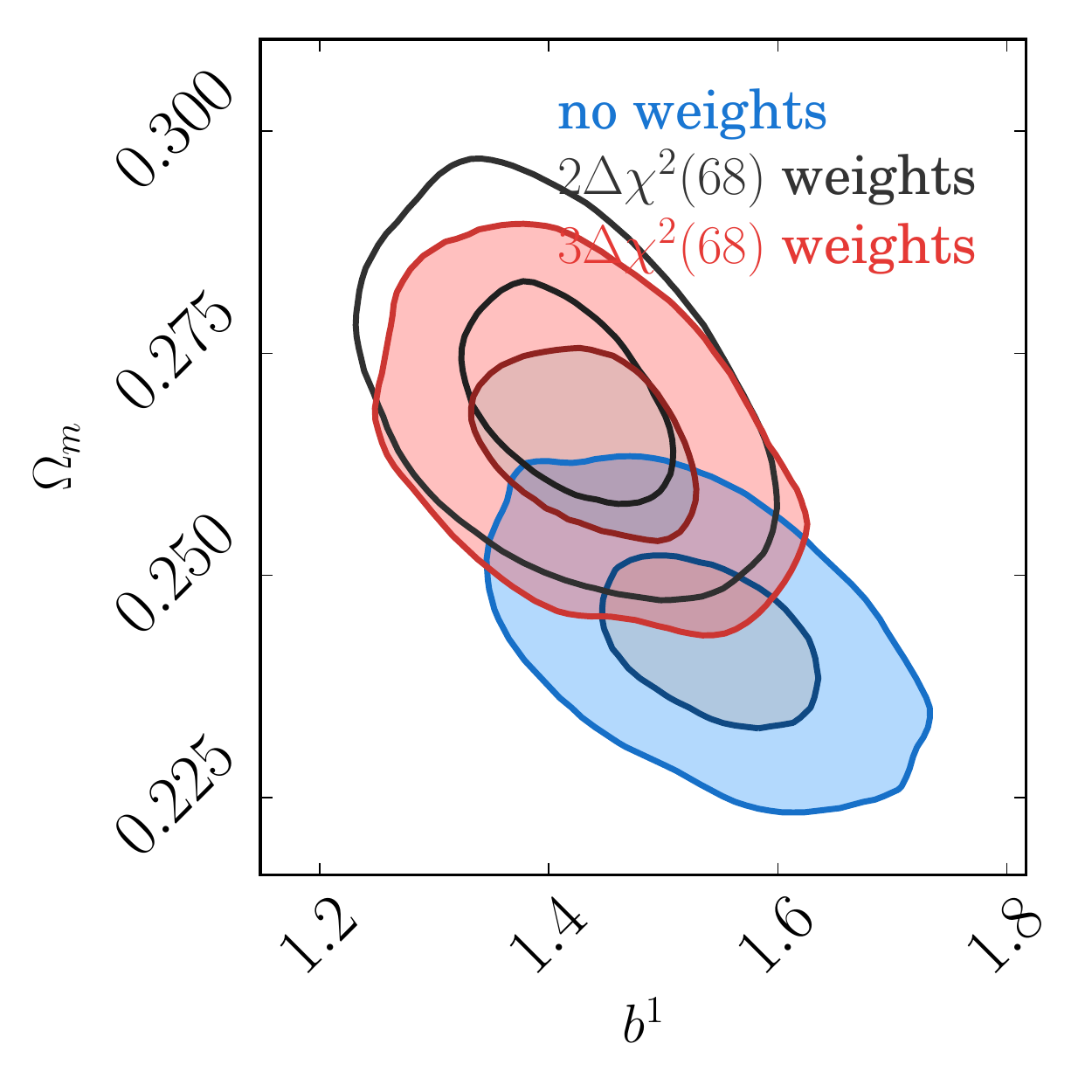}
  \includegraphics[trim= 2.5cm 0cm 0.0cm 0cm, clip, totalheight=0.3\textwidth]{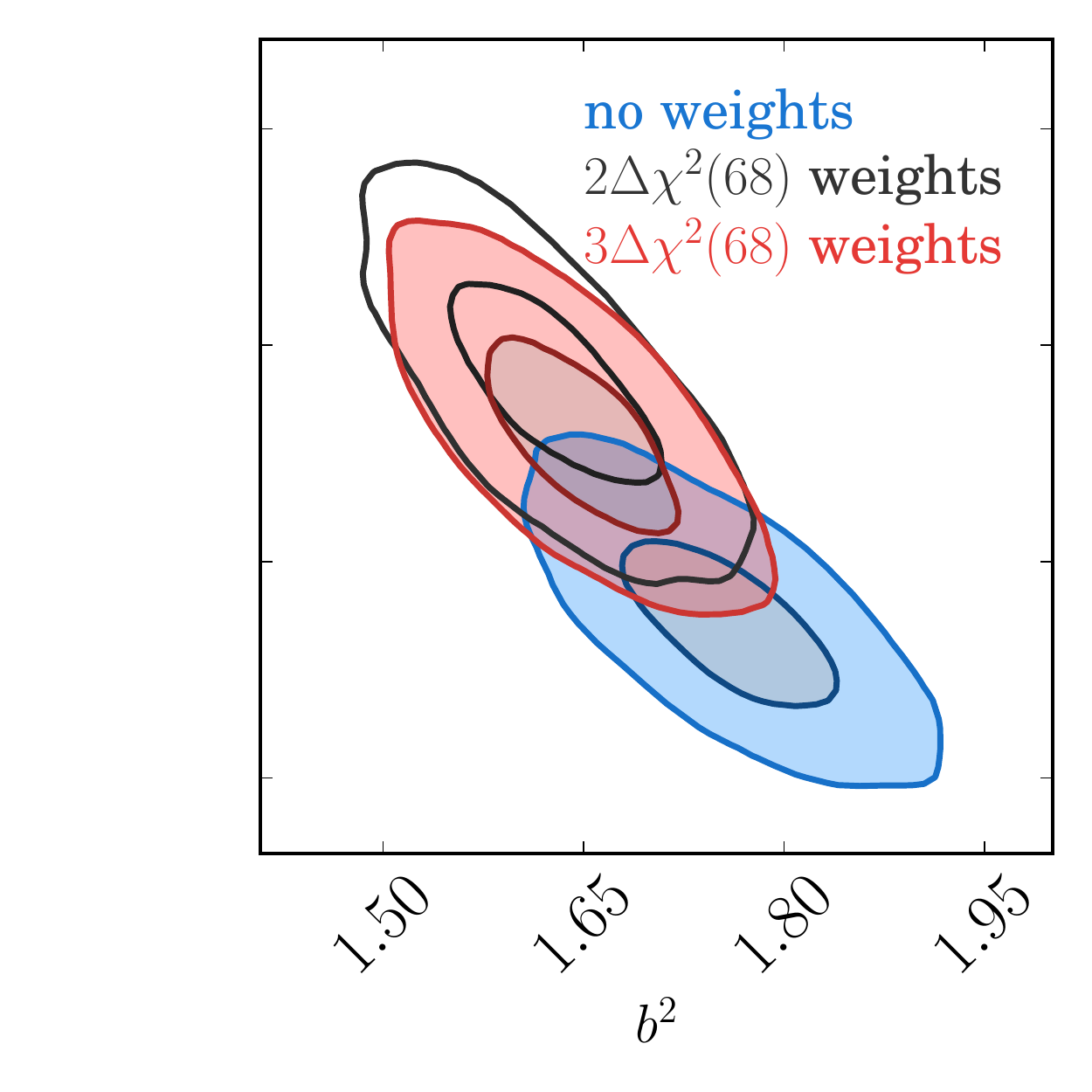}
  \includegraphics[trim= 2.5cm 0cm 0.0cm 0cm, clip, totalheight=0.3\textwidth]{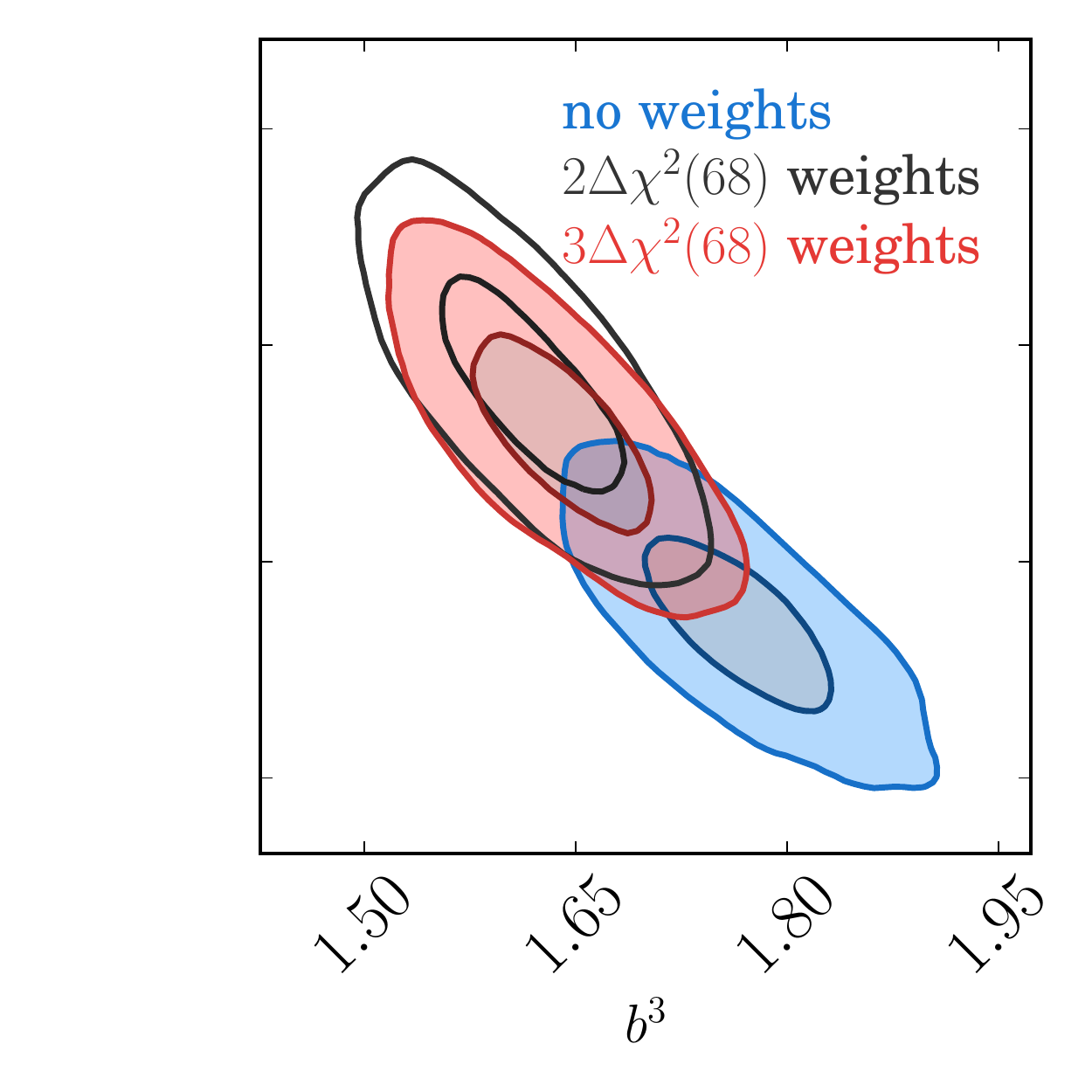}
  \includegraphics[trim= 0.0cm 0cm 0.0cm 0cm, clip, totalheight=0.3\textwidth]{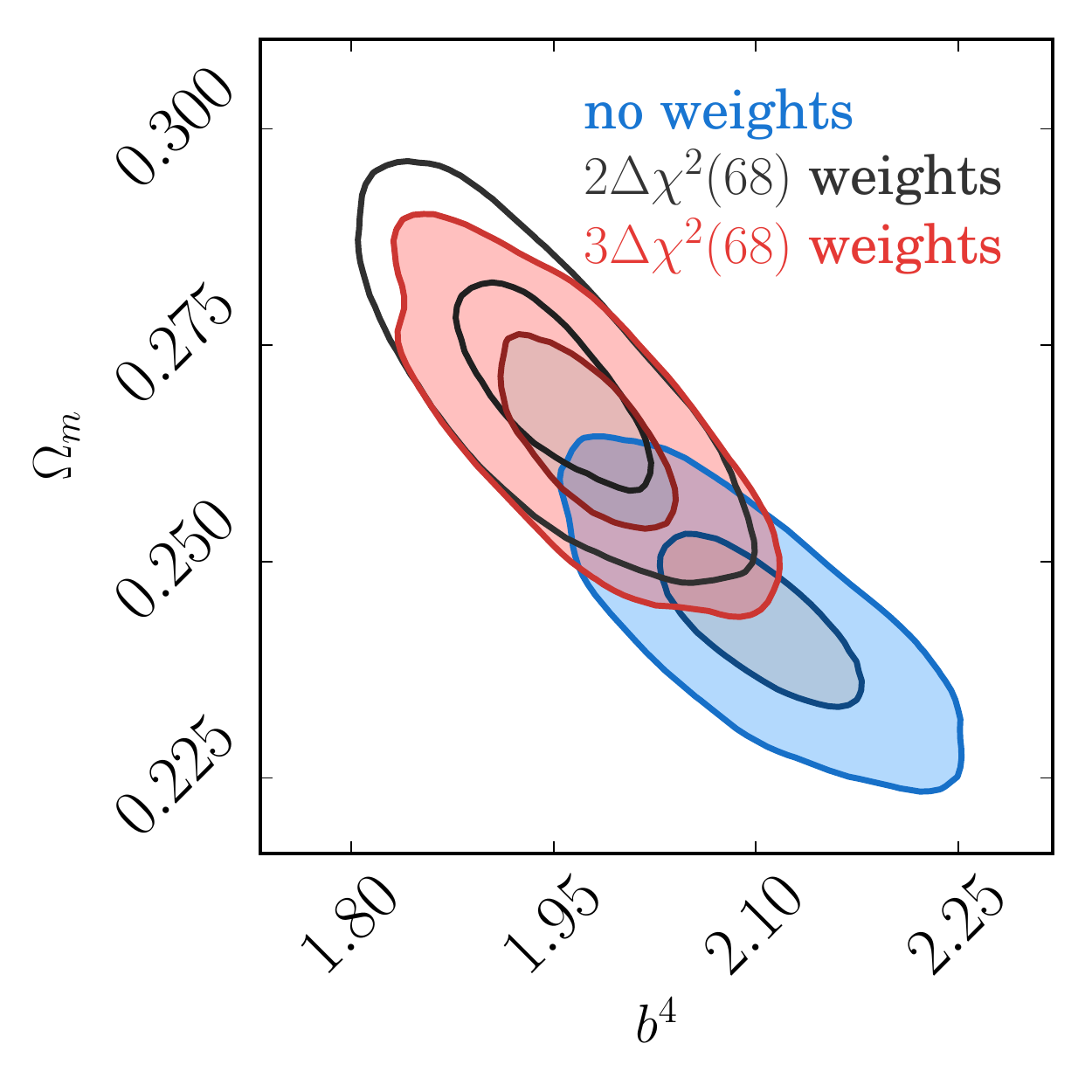}
  \includegraphics[trim= 2.5cm 0cm 0.0cm 0cm, clip, totalheight=0.3\textwidth]{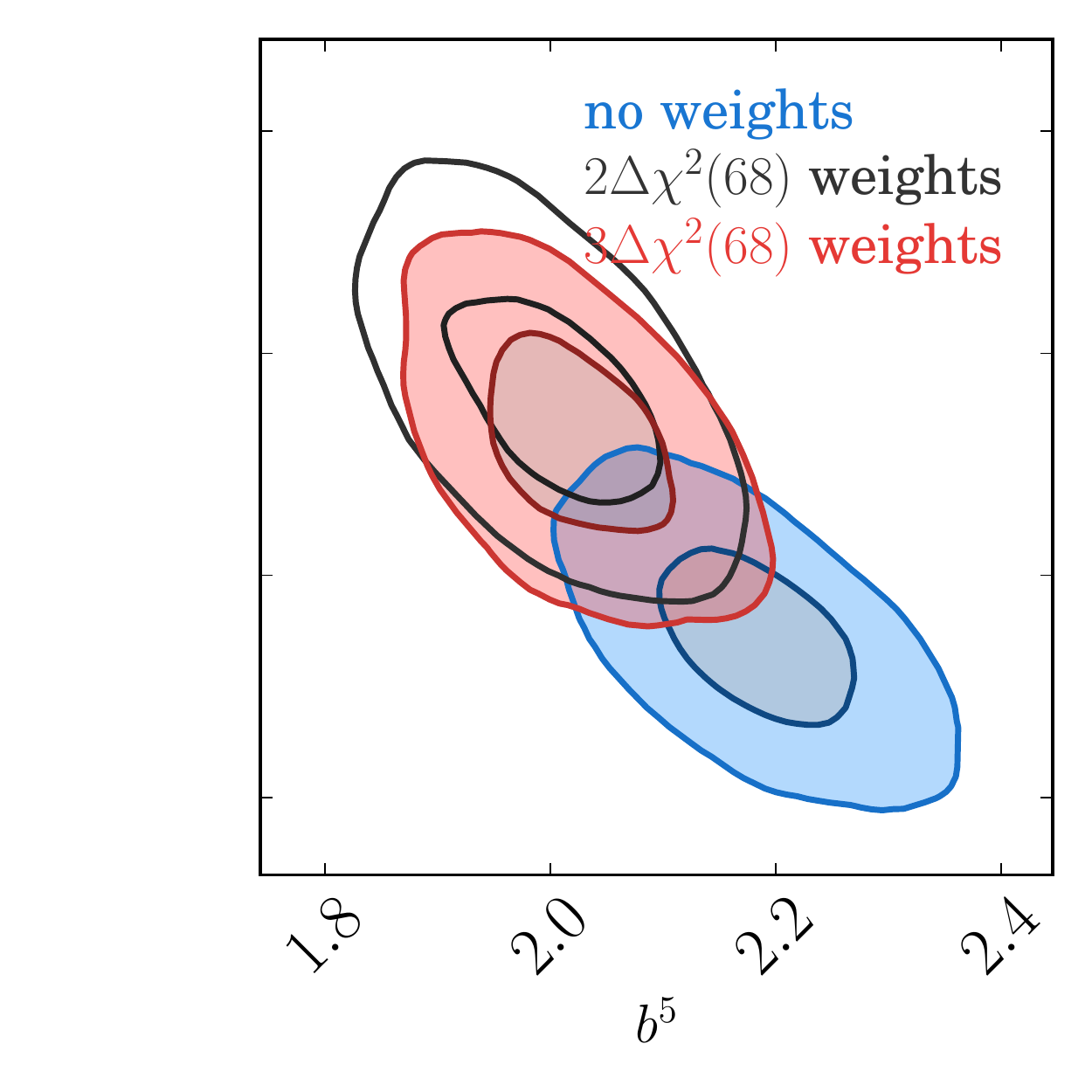}
  \caption{Parameter constraints showing the impact of the SP weights, varying $\Omega_m$, 5 linear bias parameters $b^{i}$, and 5 nuisance parameters $\Delta z^{i}$. Contours are drawn at 68\% and 95\% confidence level. These constraints use the same $\Delta z^{i}$ priors as Y1COSMO. The blue contour shows the constraints on $w(\theta)$ calculated with no SP weights. The gray and red contours use SP weights removing all $2\Delta \chi^{2}(68)$ and $3\Delta \chi^{2}(68)$ correlations respectively. In this parameter space, ignoring  the correlations with survey properties would have significantly biased the constraints from $w(\theta)$. As expected, the best fit when using the $2\Delta \chi^{2}(68)$ weights is at smaller values of $b^{i}$ than the $3\Delta \chi^{2}(68)$ weights, although the difference is not significant compared to the size of the contour.}
  \label{constraints_weights_impact}
\end{figure*}

\begin{figure*}
  \includegraphics[trim= 0.0cm 0cm 0.0cm 0cm, clip, totalheight=0.3\textwidth]{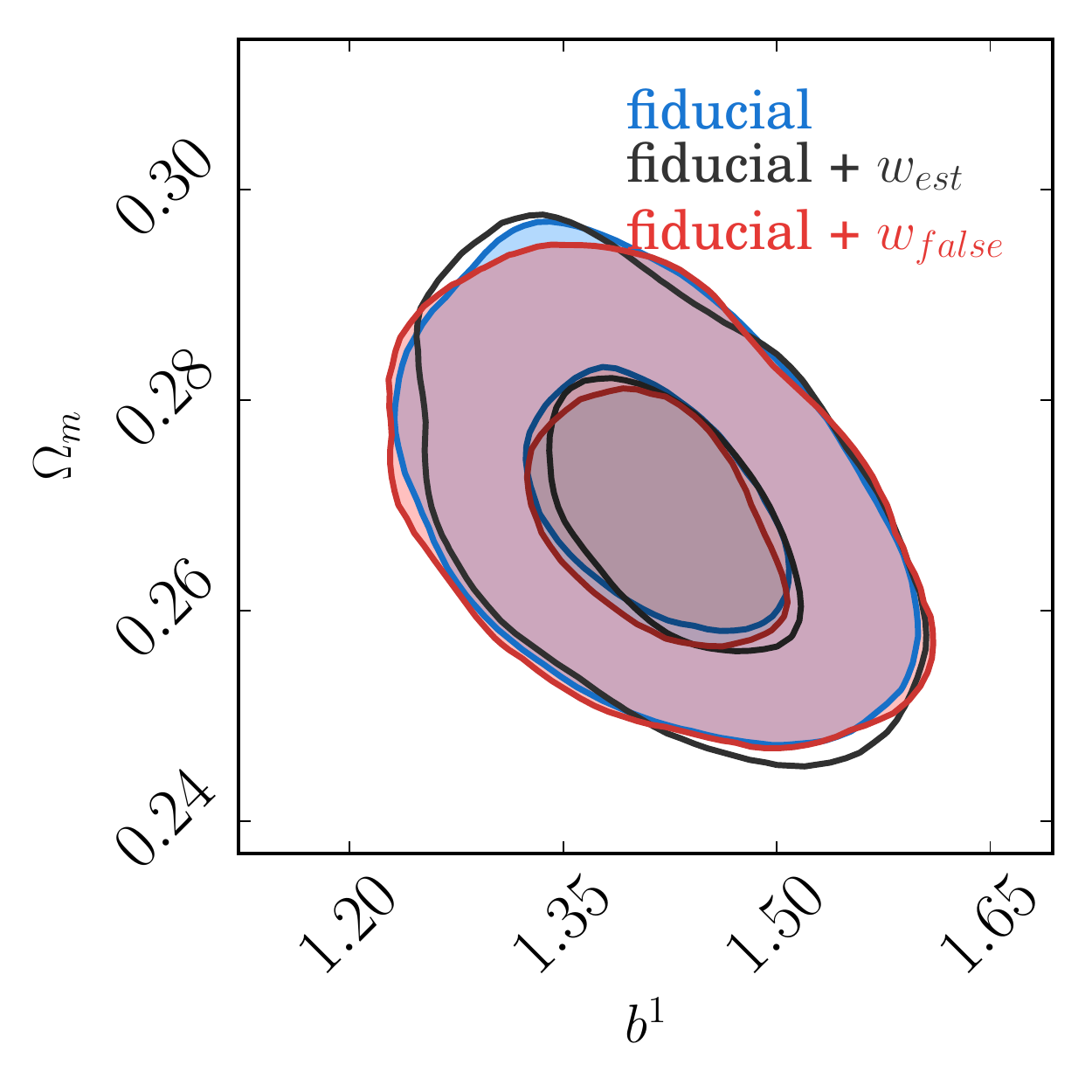}
  \includegraphics[trim= 2.5cm 0cm 0.0cm 0cm, clip, totalheight=0.3\textwidth]{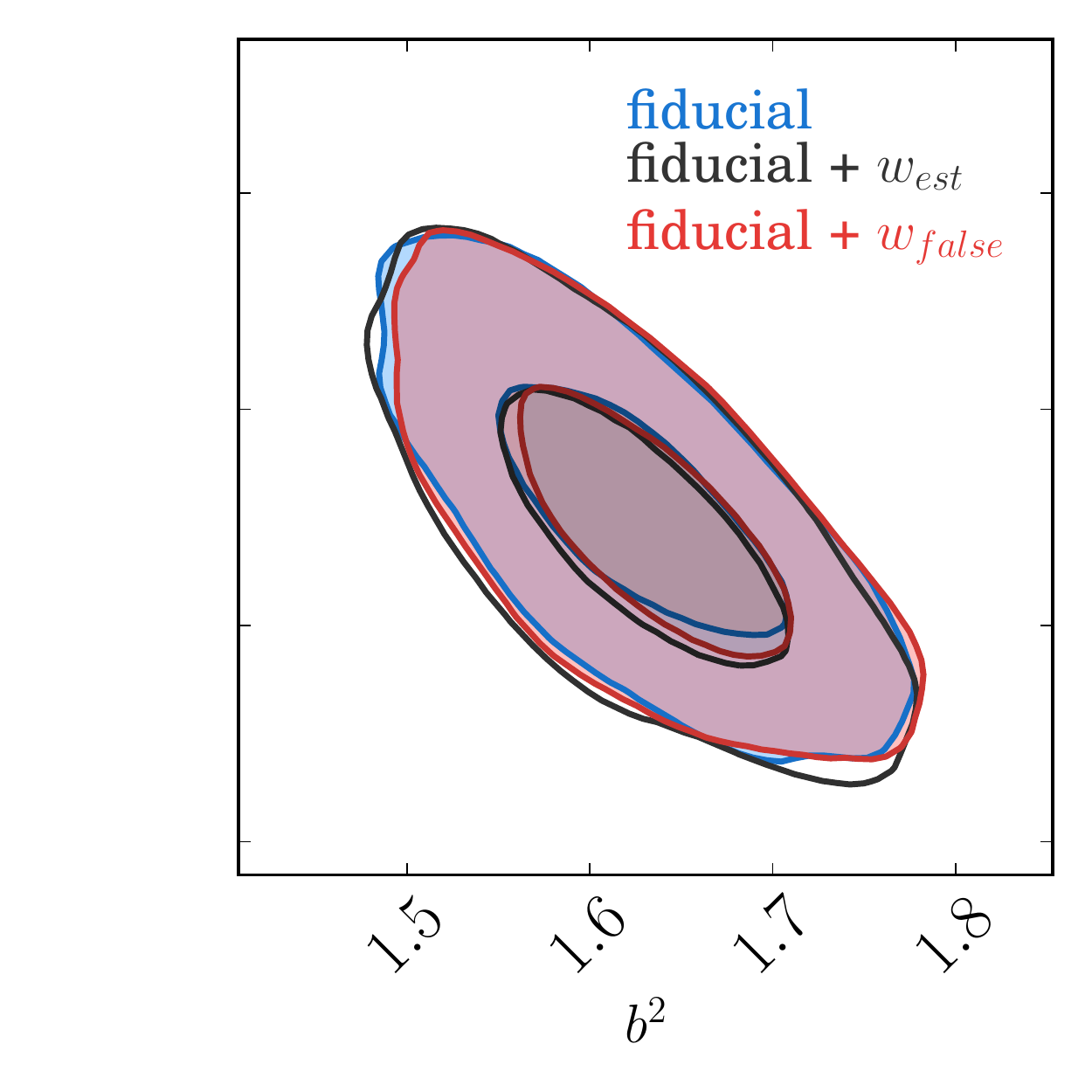}
  \includegraphics[trim= 2.5cm 0cm 0.0cm 0cm, clip, totalheight=0.3\textwidth]{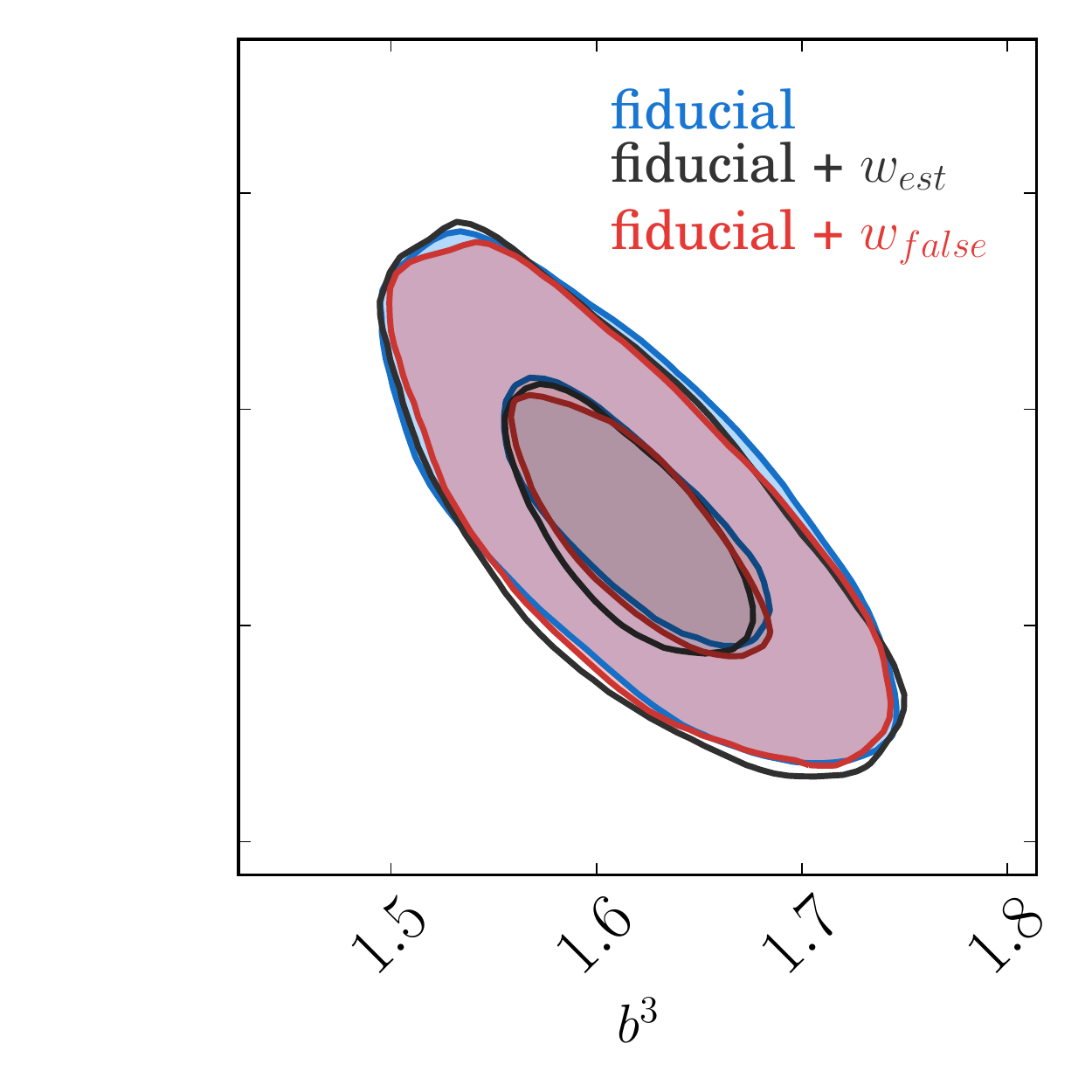}
  \includegraphics[trim= 0.0cm 0cm 0.0cm 0cm, clip, totalheight=0.3\textwidth]{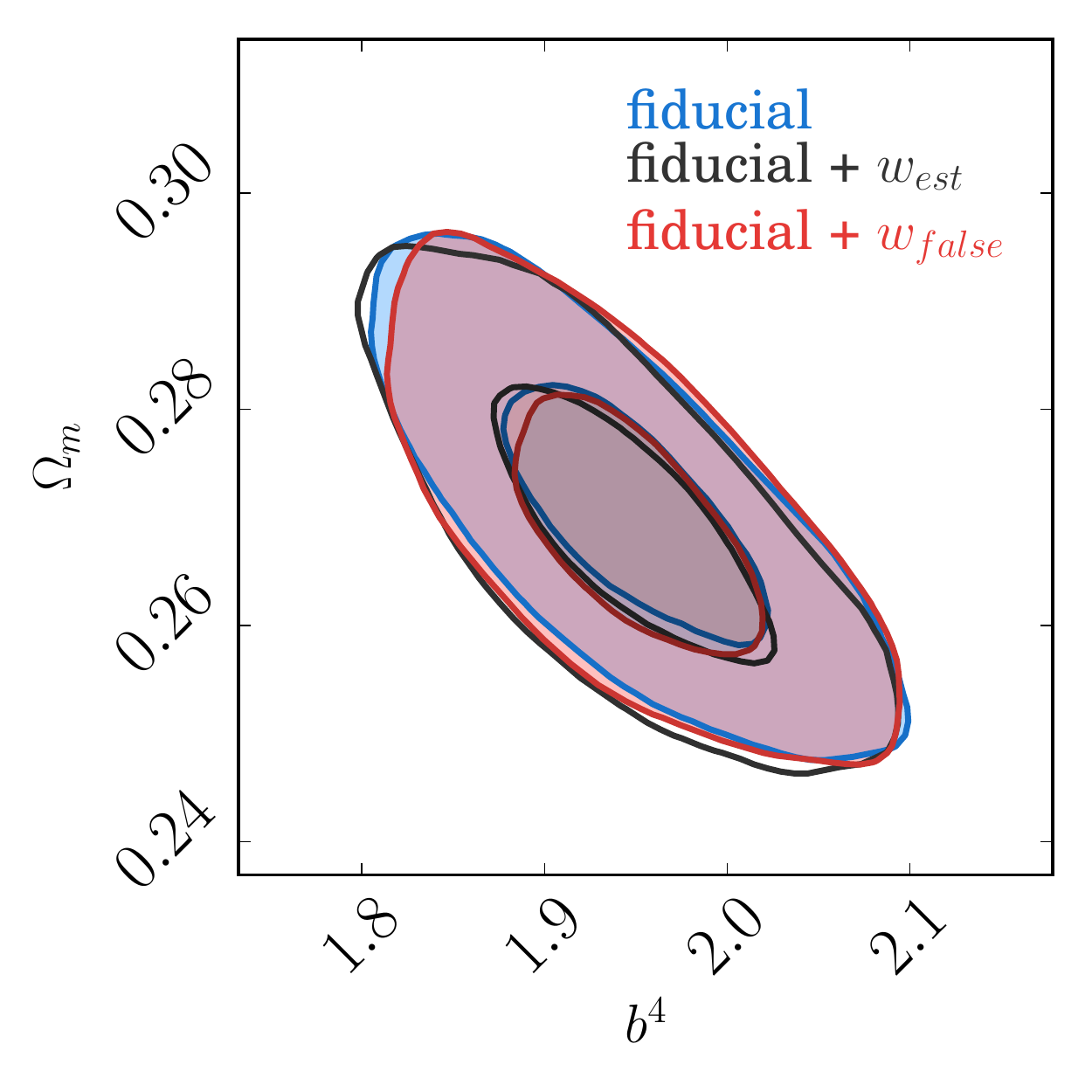}
  \includegraphics[trim= 2.5cm 0cm 0.0cm 0cm, clip, totalheight=0.3\textwidth]{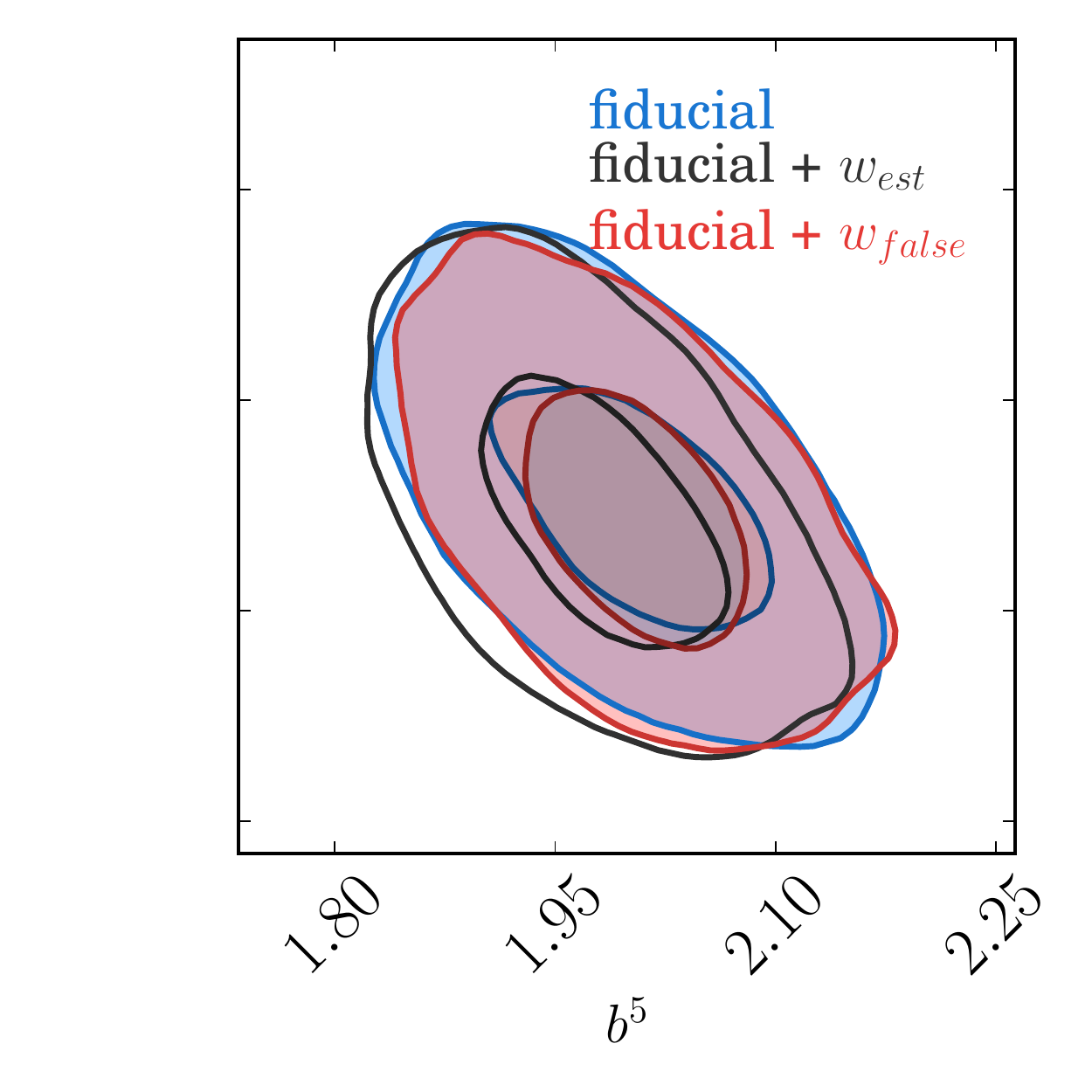}
  \caption{ Parameter constraints showing the impact of the estimator bias, $w_{\rm est}$ and false correction bias $w_{\rm false}$. The fiducial data vector and was calculated using the $2\Delta \chi^{2}(68)$ weights on the data. The $w_{\rm est}$ and $w_{\rm false}$ were measured on Gaussian mock surveys using a $2\Delta \chi^{2}(68)$ threshold significance. We see no evidence for significant bias in the $b^{i}$, $\Omega_{m}$ plane. These constraints use the same $\Delta z^{i}$ priors as Y1COSMO.} 
  \label{constraints_westbias_wfalsebias}
\end{figure*}

\begin{figure*}
  \includegraphics[trim= 0.0cm 0cm 0.0cm 0cm, clip, totalheight=0.3\textwidth]{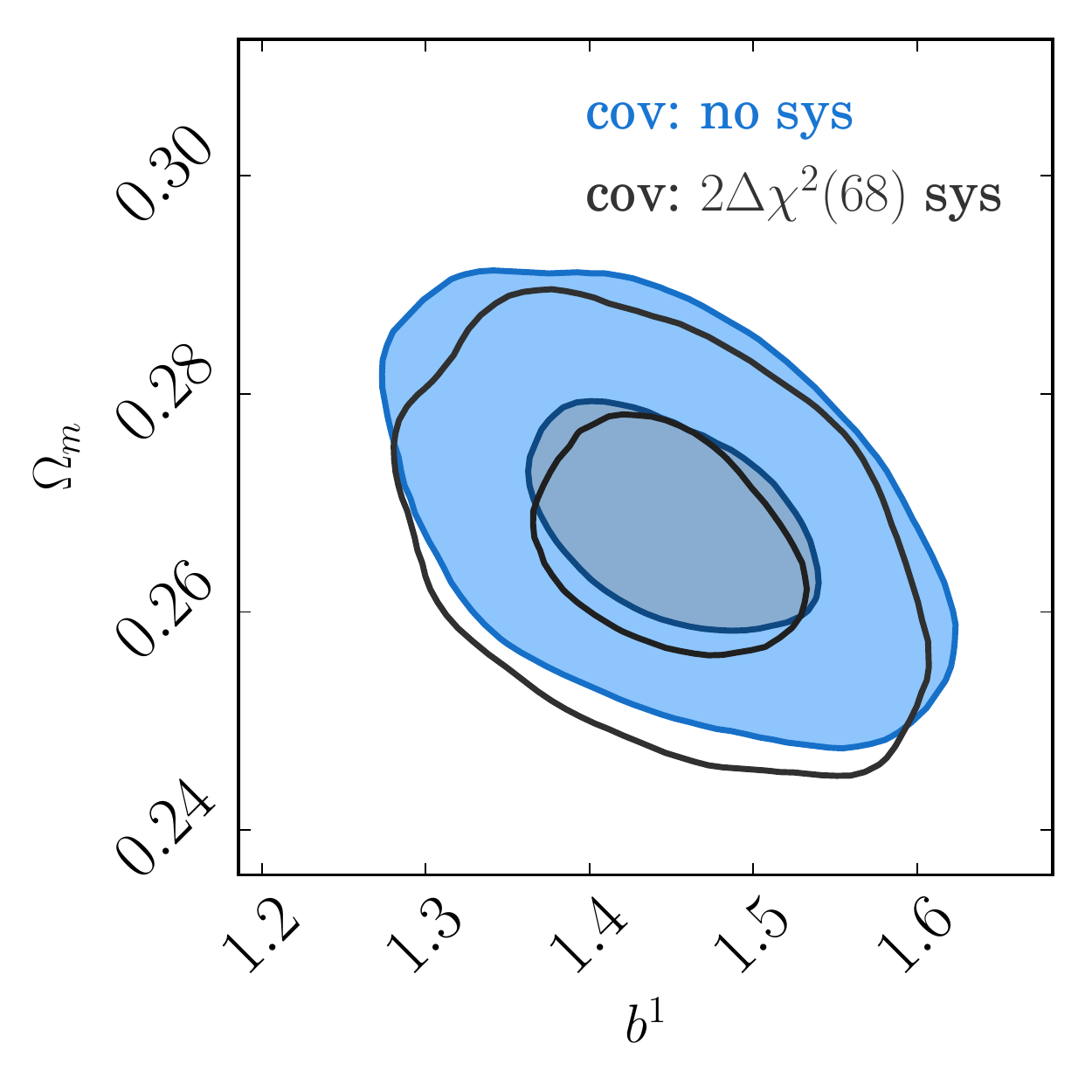}
  \includegraphics[trim= 2.5cm 0cm 0.0cm 0cm, clip, totalheight=0.3\textwidth]{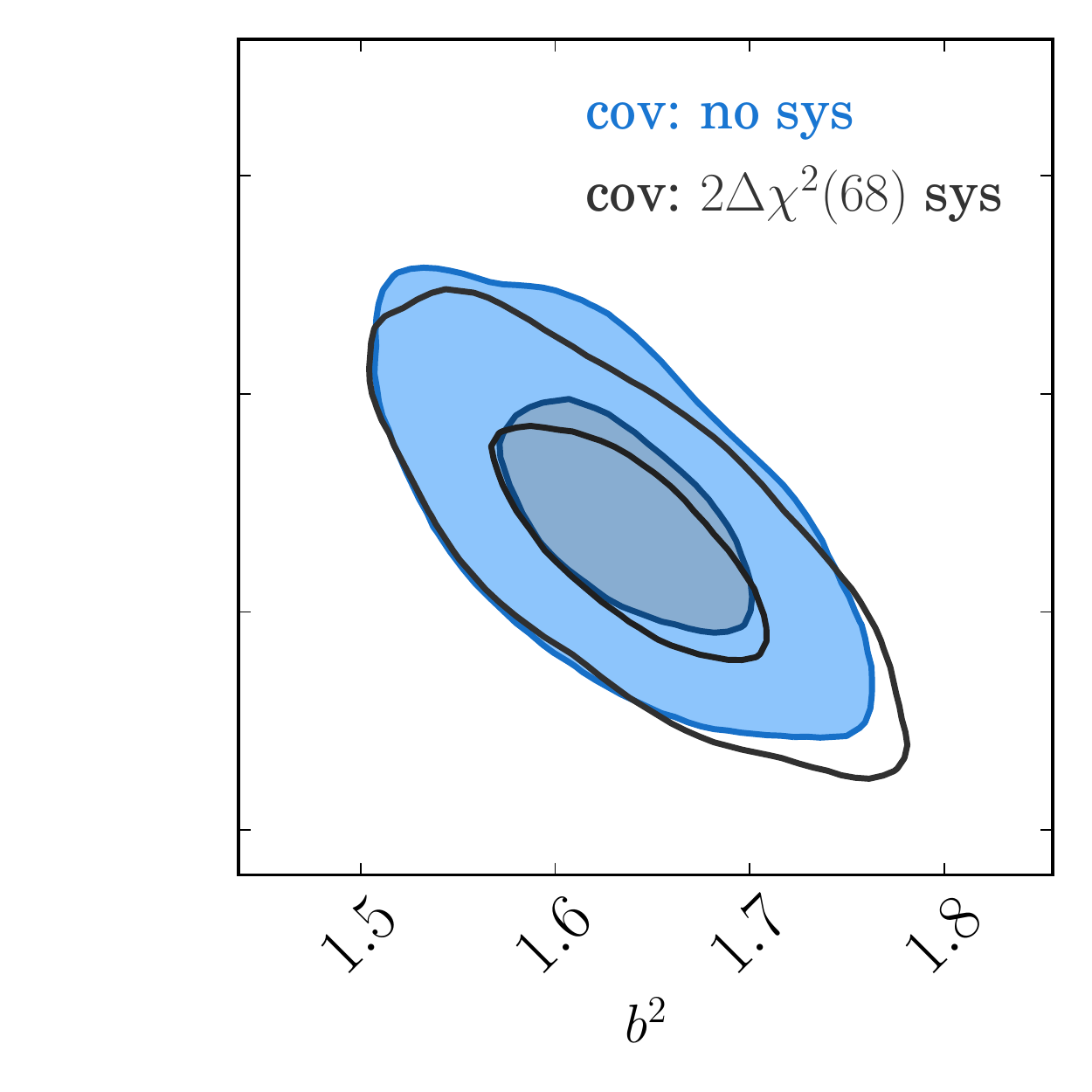}
  \includegraphics[trim= 2.5cm 0cm 0.0cm 0cm, clip, totalheight=0.3\textwidth]{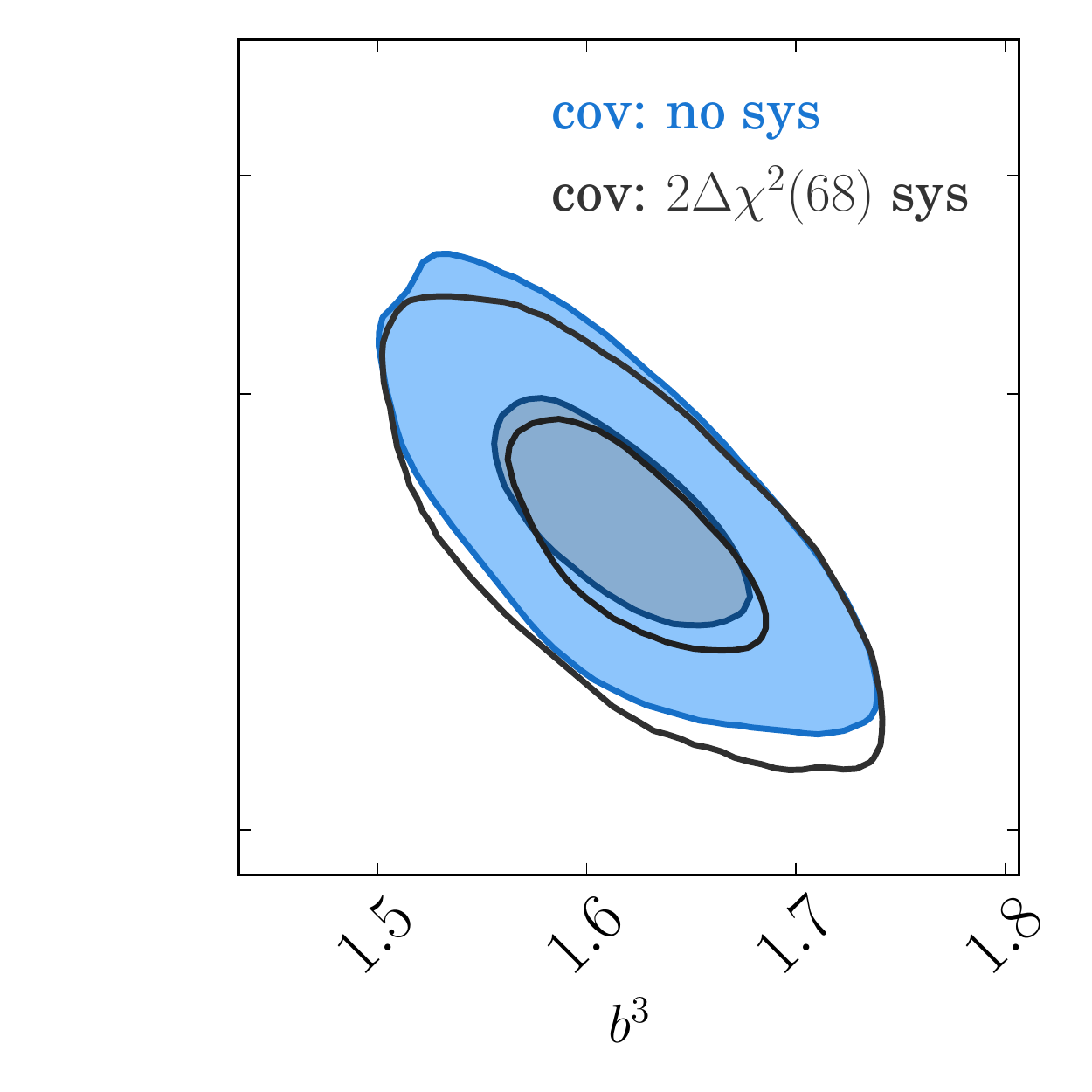}
  \includegraphics[trim= 0.0cm 0cm 0.0cm 0cm, clip, totalheight=0.3\textwidth]{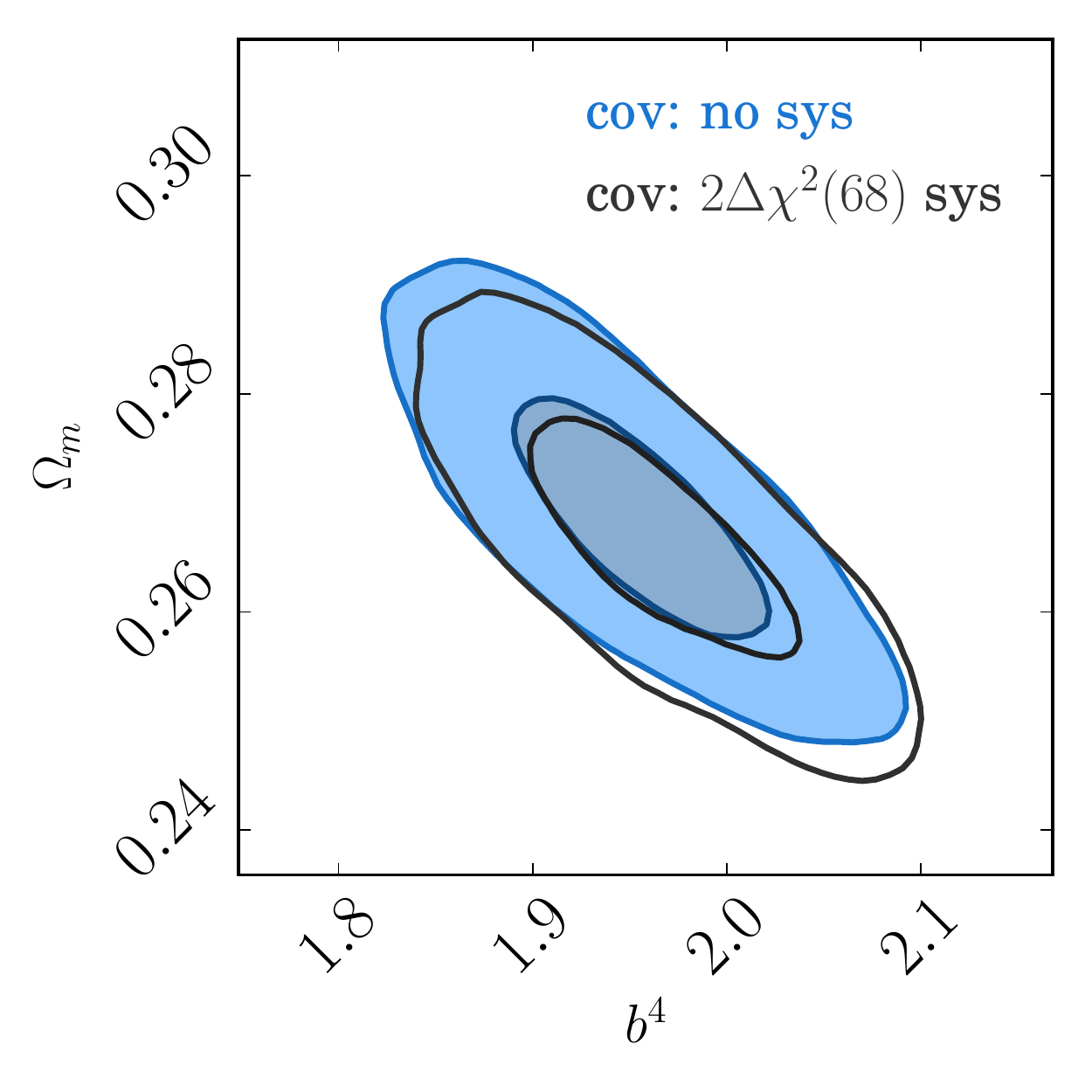}
  \includegraphics[trim= 2.5cm 0cm 0.0cm 0cm, clip, totalheight=0.3\textwidth]{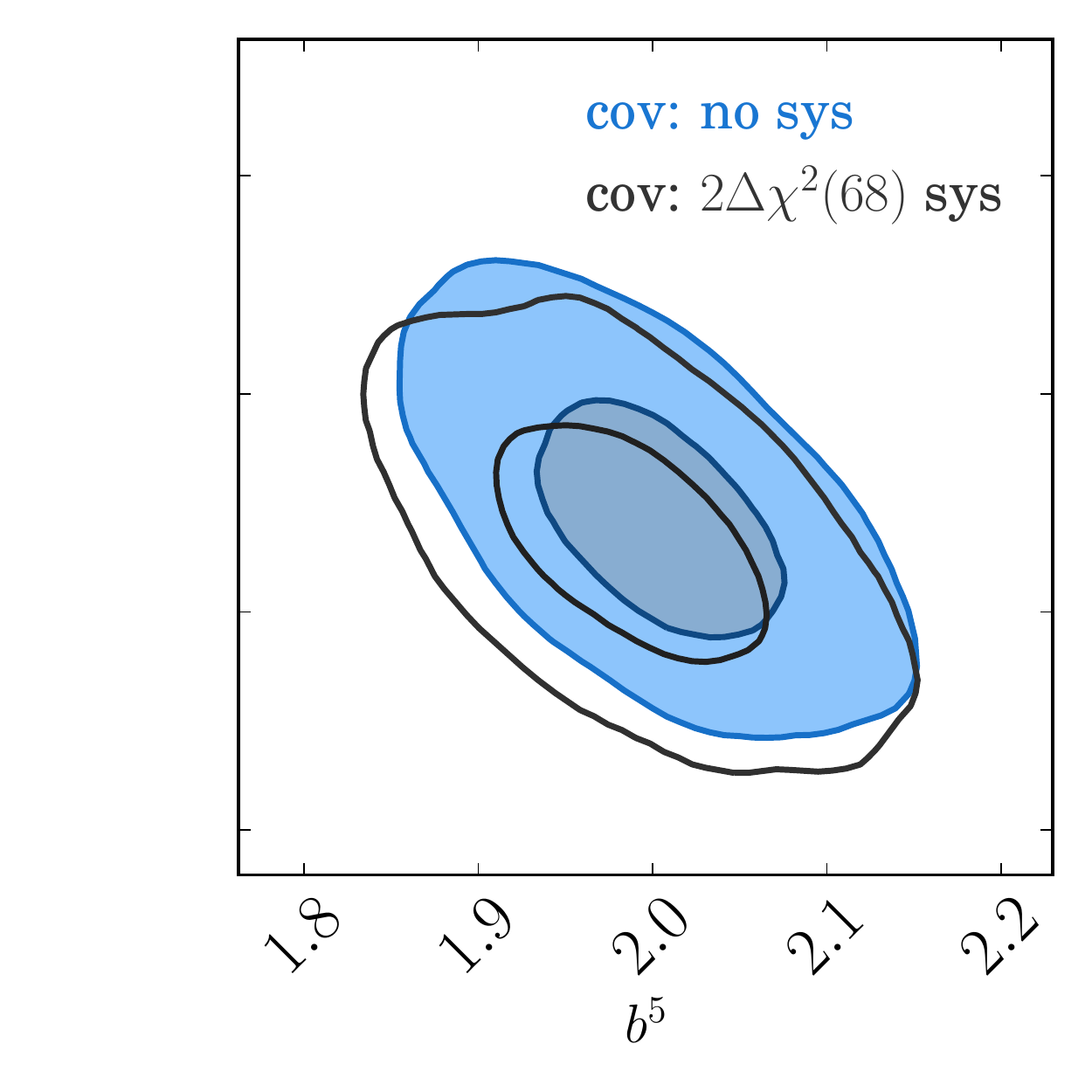}
  \caption{Parameter constraints showing the impact of the systematics correction on the covariance. Both contours use the fiducial theory data vector. The blue contour uses the covariance from mock surveys with no contamination added (labeled "cov: no sys"). The gray contour uses the covariance determined from mock surveys with the $2\Delta \chi^{2}(68)$ contaminations added (labeled "cov: $2\sigma$ sys"). These constraints use the same $\Delta z^{i}$ priors as Y1COSMO.}
  \label{constraints_cov_impact}
\end{figure*}

The angular correlation function has been calculated on scales below $\theta^{i}_{\rm min}$, but these were removed in all parameter constraints. 

Fixing all cosmological parameters, including $\Omega_m$, at the Y1COSMO values, we measure the linear bias to be $b_{1}=1.40 \pm 0.07$,  $b_{2}=1.60 \pm 0.05$,  $b_{3}=1.60 \pm 0.04$,  $b_{4}=1.93 \pm 0.04$, and $b_{5}=1.98 \pm 0.07$. The $\chi^{2}$ values of the combined fit and the individual bins are shown in Table \ref{chi2_table}. 
We note that the bin with the smallest probability is bin $1$. 

The combined goodness-of-fit $\chi^{2}$ of the bias measurements is $\chi^{2} = 67$ and the number of degrees of freedom is $\nu = 54-10$ (the 10 parameters are $b_{i}$, $\Delta z_{i}$). These values provide a probability to exceed of $1.4\%$. As in Y1COSMO, we note that the formal probabilities of a $\chi^{2}$ distribution are not strictly applicable in this case due to the uncertainty on the estimates of the covariance. Further, because the five $\Delta z_{i}$ are nuisance parameters with tight priors, we also consider $\nu = 49$, which yields a probability to exceed of $4.5\%$. These probabilities are very similar to the values obtained by the full Y1COSMO data vector, of which this is one part.

We also note that the $\chi^{2}$ is sensitive to the inclusion of the shot-noise correction applied to the covariance detailed in Y1COSMO whereas the $b^{i}$ values and uncertainty were insensitive to this change. 

\begin{table}
\begin{center}
\begin{tabular}{p{3cm}p{1.5cm}p{1.5cm}p{1.5cm}}
	\hline
	$z$ range & $\chi^{2}$ & $N_{\rm data}$ & prob\\
    \hline
	$0.15 < z < 0.3 $     & $14.8$ & $8$  & $2.2\% $ \\
	$0.3  < z < 0.45$     & $6.9$  & $10$ & $55\%  $ \\
	$0.45 < z < 0.6 $     & $17.7$ & $11$ & $3.9\% $ \\
	$0.6  < z < 0.75$     & $11.0$ & $12$ & $35.9\%$ \\
	$0.75 < z < 0.9 $     & $16.5$ & $13$ & $12.2\%$ \\
    $w(\theta)$ all bins & $67.2$ & $54$ & $1.4\%$\\
  	\hline
\end{tabular}
\end{center}
\caption{The $\chi^{2}$, and probability of obtaining a $\chi^{2}$ exceeding this values for each redshift bin and for all bins combined. For the combined $\chi^{2}$, the number of free parameters is 10 (5 $b_{i}$ and 5 $\Delta z_{i}$).
The individual $z$ bin $\chi^{2}$ values are calculated using the best fit to all $z$ bins combined. The covariance between between $z$ bins is sufficiently small that we can treat these as independent. We have therefore considered each individual bin to have 2 free parameters. It is expected that measuring the bias in each bin separately would have resulted in a smaller $\chi^{2}$. 
}
\label{chi2_table}
\end{table}

For the $L/L_*>0.5$ sample, the bias is nearly constant as a function of redshift, though there is a decrease at low redshift that has more than 2$\sigma$ significance (the correlation in the measured bias for bins 1 and 3 is only -0.04, so we can safely ignore it in this discussion). The difference between bin 1 and bin 3 is less significant if we determine the expectation for a passively evolving sample as in \cite{Fry96,Tegmark96bias}, which predicts a bias of 1.52 at $z=0.24$ given a bias of 1.61 at $z=0.53$. The bias increases for the higher luminosity sample, as expected. The results are broadly consistent with previous studies of the bias of red galaxies at low redshift (see, e.g., \cite{Ross11red} for a review) and BOSS at intermediate redshifts (see, e.g., \cite{Saito16}). Further study of the details of the \textsc{redMaGiC} samples is warranted, especially if one wishes to use $w(\theta)$ at scales smaller than those studied in Y1COSMO.

We compare these bias constraints to those measured from the galaxy-galaxy lensing probe of the same \textsc{redMaGiC} sample, presented in Y1GGL. We parameterize the difference between the two measurements with the cross-correlation coefficients $r^{i}$, which are presented in Figure \ref{r_plot}. Beyond linear galaxy bias, $r$ can deviate from 1 and acquire scale dependences, and it must be properly modeled to constrain cosmology with combined galaxy clustering and galaxy-galaxy lensing (e.g.\ \cite{baldauf+09}). We constrain $r^{i}$ at fixed cosmology using the Y1COSMO covariance, which includes the covariance between the two probes. All the nuisance parameters discussed in Y1COSMO are varied for this constraint. With our choice of scale cuts, we see no evidence of tension between the two bias measurements. This provides further justification for fixing $r = 1$ in the Y1COSMO analysis.

\section{Demonstration of Robustness}
\label{sec:robust}
We apply a number of null tests to our weighted sample to demonstrate its robustness. We do so by obtaining constraints on the galaxy bias and $\Omega_m$. These parameters are sensitive to both multiplicative and additive shifts in the amplitude of $w(\theta)$ and we therefore believe they should encapsulate any potential systematic bias that could affect the cosmological analysis of Y1COSMO. We thus perform joint fits to the data in each redshift bin to obtain constraints on the five $b^{i}$ and $\Omega_m$. For these fits, we marginalize over an additive redshift bias uncertainty described in Table \ref{bin_table2}. All other cosmological parameters are fixed at the Y1COSMO cosmology and as such, this should not be interpreted as a measurement of $\Omega_m$ to be used in further analyses. Results are obtained using the analysis pipeline described in \cite{MPPmethodology}. We describe how $w(\theta)$ is altered to perform each test throughout the rest of this section.

\subsection{Selection of threshold}
\label{sec:threshold}

We test two thresholds used to determine when to apply weights based on a given SP map: 3$\Delta \chi^{2}(68)$ and a more restrictive (i.e., more maps weighted for) $2\Delta \chi^{2}(68)$. After reaching a certain threshold, we expect that the only effect from adding extra weights would be to bias the measurements (from over-correction) and add greater uncertainty. We test for those effects in the following subsections. Here, in order to demonstrate that our results are insensitive to the choice in threshold, the change in the measured $b^{i}$ and $\Omega_m$ must be negligible compared to its uncertainty.

Figure \ref{constraints_weights_impact} shows the difference between the $3\Delta \chi^{2}(68)$ and $2\Delta \chi^{2}(68)$ SP weights. Because the weights correction can only decrease the $w(\theta)$ signal, applying a stricter threshold significance is expected to move the contours towards smaller values of $b^{i}$. Figure \ref{constraints_weights_impact} shows that this impact is very small compared to the overall Y1 uncertainty and we can conclude that the choice between $3\Delta \chi^{2}(68)$ and $2\Delta \chi^{2}(68)$ weights will have negligible impact on the Y1COSMO parameter constraints (The final weights used in Y1COSMO are the $2\Delta \chi^{2}(68)$ weights). 

Figure \ref{constraints_weights_impact} also shows the impact of not including SP weights on the parameter constraints. Ignoring the SP correlations would have resulted in significantly biased constraints on $b^{i}$ and $\Omega_m$. In every redshift bin, the shift is greater than 2$\sigma$ along the major axis of the ellipses.

\subsection{Estimator bias}
\label{estimatorbias}

We also test for potential bias in $w(\theta)$ induced by over-correcting with the weights method and from correlations between the SP maps. This was done using the Gaussian mocks described in Section \ref{sec:covariances} using the following method. After the galaxy over-density field has been generated in each realization, we insert the systematic correlation using $F_{\rm sys}(s)$ and the best-fit parameters for each of the systematics in Table \ref{weights_table} at 2$\Delta\chi^2(68)$ significance. This is equivalent to dividing each mock galaxy map by a map of the SP weights. 

We then produced a galaxy number count as before, also adding shot noise. We fit the parameters of $F_{\rm sys}(s)$ to each realization and apply weights to the maps using the same method that is applied to the data. We measure $w(\theta)$ using the pixel estimator in Eqn. \ref{eqn:pixel_estimator} on mocks with systematic contamination and correction, $w_{\rm weights}$, and on mocks with no systematics added, $w_{\rm no \ sys}$. We define the bias in $w(\theta)$ to be,

{\setlength{\mathindent}{0cm}
\begin{equation}
w_{\rm est \ bias} = \frac{1}{N} \left( \sum_{i=1}^{N} w_{{\rm no \ sys},i} - \sum_{j=1}^{N} w_{{\rm weights},j} \right)
\end{equation}}

\noindent where $N$ is the total number of realizations. We then add $w_{\rm est \ bias}$ to the measured $w(\theta)$ and measure $b^{i}$ and $\Omega_{m}$. This is designed to test for any bias in $w(\theta)$ induced by the the estimator when using weights. 

This result can be seen in Figure \ref{constraints_westbias_wfalsebias} where it shows negligible impact on the parameter constraints.

\subsection{False correlations}
\label{falsebias}

Given the large number of SP maps being used in the systematics tests, it is possible that chance correlations will appear significant and weights will be applied where no contamination has occurred, biasing the measured signal. To test this, we use the same Gaussian mocks as in Section \ref{estimatorbias} with no added systematic contaminations. We measure the correlation of each mock with each of the 21 SP maps in Section \ref{sp_maps}, identifying any correlations above a $2\Delta \chi^{2}(68)$ threshold significance. 

The false correction bias $w_{\rm false \ bias}$, is then defined as the average difference between the $w(\theta)$ measured with no corrections, and the $w(\theta)$ measured correcting for all correlations above the threshold using the weights method. We then add $w_{\rm false \ bias}$ to the measured $w(\theta)$ and test the impact on $b^{i}$ and $\Omega_m$ constraints.

This test is designed to test for any bias in $w(\theta)$ induced by falsely correcting for SP maps that were only correlated with the galaxy density by chance. 

This result is shown in Figure \ref{constraints_westbias_wfalsebias} where $w_{\rm false \ bias}$ for the $2\Delta \chi^{2}(68)$ SP maps has been used. This shows a negligible impact on the constraints. The $w_{\rm false \ bias}$ for the $3\Delta \chi^{2}(68)$ SP maps is not shown as it has an even smaller impact.  This demonstrates that selecting a $2\Delta \chi^{2}(68)$ threshold does not induce a bias in the inferred bias parameters for the set of SP maps used in this analysis.

\subsection{Impact on covariance}

Correcting for multiple systematic correlations can alter the covariance of the $w(\theta)$ measurement in various ways. We expect that scatter in the best fit parameters should increase the variance, while the removal of some clustering modes should decrease it. We test the significance of any changes to the amplitude and structure of the covariance matrix using the Gaussian mocks. 

For this test we use the same mocks as in Section \ref{estimatorbias} which are `contaminated' with the same systematic correlations found in the data. We fit the $F_{\rm sys}(s)$ function to each mock and correct using weights. We then measure the correlation function $w_{\rm weights}$ and calculate the covariance matrix of this measurement. We also measure the correlation on mocks with no systematics added, $w_{\rm no \ sys}$, and calculate the covariance matrix from each measurement. We calculate the galaxy bias $b^{i}$ and $\Omega_{m}$ constraints for each covariance matrix and test if the resulting contours are significantly different. This test determines whether this additional uncertainty needs to be considered in the Y1COSMO analysis by marginalizing over the fitted parameters. 

The results of this test are shown in Figure \ref{constraints_cov_impact}.  We show that for the SP maps selected in this analysis, the impact on the size of the contours is negligible. We have therefore not included any additional parameters in the MCMC analysis to account for the uncertainty in the correction. 

\section{Conclusions}
\label{sec:conclusions}

We have presented the 2-point angular galaxy correlation functions, $w(\theta)$, for a sample of luminous red galaxies in DES Y1 data, selected by the \textsc{redMaGiC} algorithm. This yielded a sample with small redshift uncertainty, a wide redshift range, and wide angular area. We split this sample into five redshift bins and analyzed its clustering. Our findings can be summarized as follows:

$\bullet$ We find that multiple systematic dependencies between \textsc{redMaGiC} galaxy density and survey properties must be corrected for in order to obtain unbiased clustering measurements. We correct for these dependencies by adding weights to the galaxies, following \cite{Ross12,Ross17}. 

$\bullet$ We demonstrate both that our methods sufficiently remove systematic contamination (no significant differences are found between applying a 2$\Delta \chi^{2}(68)$ and 3$\Delta \chi^{2}(68)$ threshold; see Fig. \ref{constraints_weights_impact}) and that any bias resulting from our method removing true clustering modes is insignificant (see Fig. \ref{constraints_westbias_wfalsebias}). We further demonstrate that our weighting method imparts negligible changes to the covariance matrix (see Fig. \ref{constraints_cov_impact}).

$\bullet$ We find the redshift and luminosity dependence of the bias of \textsc{redMaGiC} galaxies to be broadly consistent with expectations for red galaxies.

$\bullet$ We find that the large-scale galaxy bias is consistent with that determined by the Y1GGL galaxy-galaxy lensing measurements. This is consistent with $r=1$ at linear scales, in agreement with basic galaxy formation theory, and a key assumption in the Y1COSMO analysis. (See Fig. \ref{r_plot}.)

$\bullet$ Our results give an unbiased $w(\theta)$ data vector to be provided to the Y1COSMO analysis, and other DES year 1 combined proebes analyses.

The methods we have presented, both correcting for systematic dependencies and ensuring the robustness of these corrections, can be used as a guide for future analyses. Possible improvements to the work include incorporating image simulations \cite{balrog} and using mode projection techniques \cite{Leistedt13}.

Our galaxy bias results can be extended to study luminosity dependence within redshift bins and to use smaller scale clustering in order to determine the HOD of \textsc{redMaGiC} galaxies. Already, our bias measurements can be used to inform simulations (e.g., for the support of DES Y3 analyses) and additional HOD information would be of further benefit.

Finally, the results presented here have been optimized for combination with other cosmological probes in Y1COSMO and our work has ensured the galaxy clustering measurements do not bias the Y1COSMO results. The analysis followed a strict blinding procedure and has yielded cosmological constraints when combined with the other 2-point functions. 


\section*{Acknowledgements}

Figures \ref{constraints_weights_impact} to \ref{constraints_cov_impact} in this paper were produced with \texttt{chainconsumer} \cite{chaincomsumer}. 

Funding for the DES Projects has been provided by the U.S. Department of Energy, the U.S. National Science Foundation, the Ministry of Science and Education of Spain, 
the Science and Technology Facilities Council of the United Kingdom, the Higher Education Funding Council for England, the National Center for Supercomputing 
Applications at the University of Illinois at Urbana-Champaign, the Kavli Institute of Cosmological Physics at the University of Chicago, 
the Center for Cosmology and Astro-Particle Physics at the Ohio State University,
the Mitchell Institute for Fundamental Physics and Astronomy at Texas A\&M University, Financiadora de Estudos e Projetos, 
Funda{\c c}{\~a}o Carlos Chagas Filho de Amparo {\`a} Pesquisa do Estado do Rio de Janeiro, Conselho Nacional de Desenvolvimento Cient{\'i}fico e Tecnol{\'o}gico and 
the Minist{\'e}rio da Ci{\^e}ncia, Tecnologia e Inova{\c c}{\~a}o, the Deutsche Forschungsgemeinschaft and the Collaborating Institutions in the Dark Energy Survey. 

The Collaborating Institutions are Argonne National Laboratory, the University of California at Santa Cruz, the University of Cambridge, Centro de Investigaciones Energ{\'e}ticas, 
Medioambientales y Tecnol{\'o}gicas-Madrid, the University of Chicago, University College London, the DES-Brazil Consortium, the University of Edinburgh, 
the Eidgen{\"o}ssische Technische Hochschule (ETH) Z{\"u}rich, 
Fermi National Accelerator Laboratory, the University of Illinois at Urbana-Champaign, the Institut de Ci{\`e}ncies de l'Espai (IEEC/CSIC), 
the Institut de F{\'i}sica d'Altes Energies, Lawrence Berkeley National Laboratory, the Ludwig-Maximilians Universit{\"a}t M{\"u}nchen and the associated Excellence Cluster Universe, 
the University of Michigan, the National Optical Astronomy Observatory, the University of Nottingham, The Ohio State University, the University of Pennsylvania, the University of Portsmouth, 
SLAC National Accelerator Laboratory, Stanford University, the University of Sussex, Texas A\&M University, and the OzDES Membership Consortium.

Based in part on observations at Cerro Tololo Inter-American Observatory, National Optical Astronomy Observatory, which is operated by the Association of 
Universities for Research in Astronomy (AURA) under a cooperative agreement with the National Science Foundation.

The DES data management system is supported by the National Science Foundation under Grant Numbers AST-1138766 and AST-1536171.
The DES participants from Spanish institutions are partially supported by MINECO under grants AYA2015-71825, ESP2015-88861, FPA2015-68048, SEV-2012-0234, SEV-2016-0597, and MDM-2015-0509, 
some of which include ERDF funds from the European Union. IFAE is partially funded by the CERCA program of the Generalitat de Catalunya.
Research leading to these results has received funding from the European Research
Council under the European Union's Seventh Framework Program (FP7/2007-2013) including ERC grant agreements 240672, 291329, and 306478.
We  acknowledge support from the Australian Research Council Centre of Excellence for All-sky Astrophysics (CAASTRO), through project number CE110001020.

This manuscript has been authored by Fermi Research Alliance, LLC under Contract No. DE-AC02-07CH11359 with the U.S. Department of Energy, Office of Science, Office of High Energy Physics. The United States Government retains and the publisher, by accepting the article for publication, acknowledges that the United States Government retains a non-exclusive, paid-up, irrevocable, world-wide license to publish or reproduce the published form of this manuscript, or allow others to do so, for United States Government purposes.

MC has been funded by AYA2013-44327-P, AYA2015-71825-P and acknowledges support from the Ramon y Cajal MICINN program.  ER acknowledges support by the DOE Early Career Program, DOE grant DE-SC0015975, and the Sloan Foundation, grant FG-2016-6443. NB acknowledges the use of University of Florida’s supercomputer HiPerGator 2.0 as well as thanks the University of Florida’s Research Computing staff. JB acknowledges support from the Swiss National Science Foundation.



\bibliographystyle{apsrev4-1}
\bibliography{main} 




\appendix

\section{Survey Property Maps}
\label{appendix:sys}

\counterwithin{figure}{section}
\setcounter{figure}{0}

\begin{figure*}
  \centering
  \includegraphics[scale=0.5]{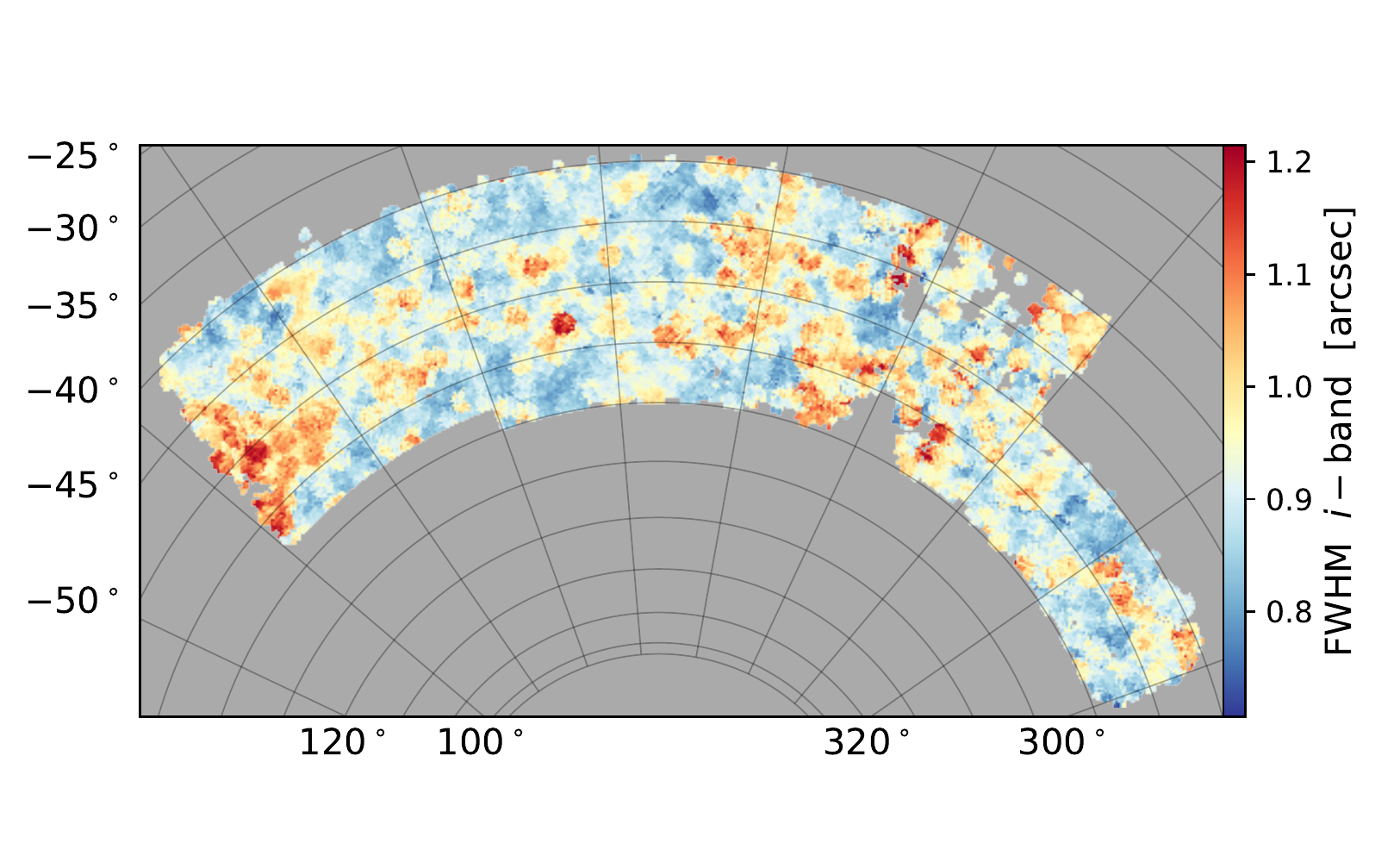}
  \includegraphics[scale=0.5]{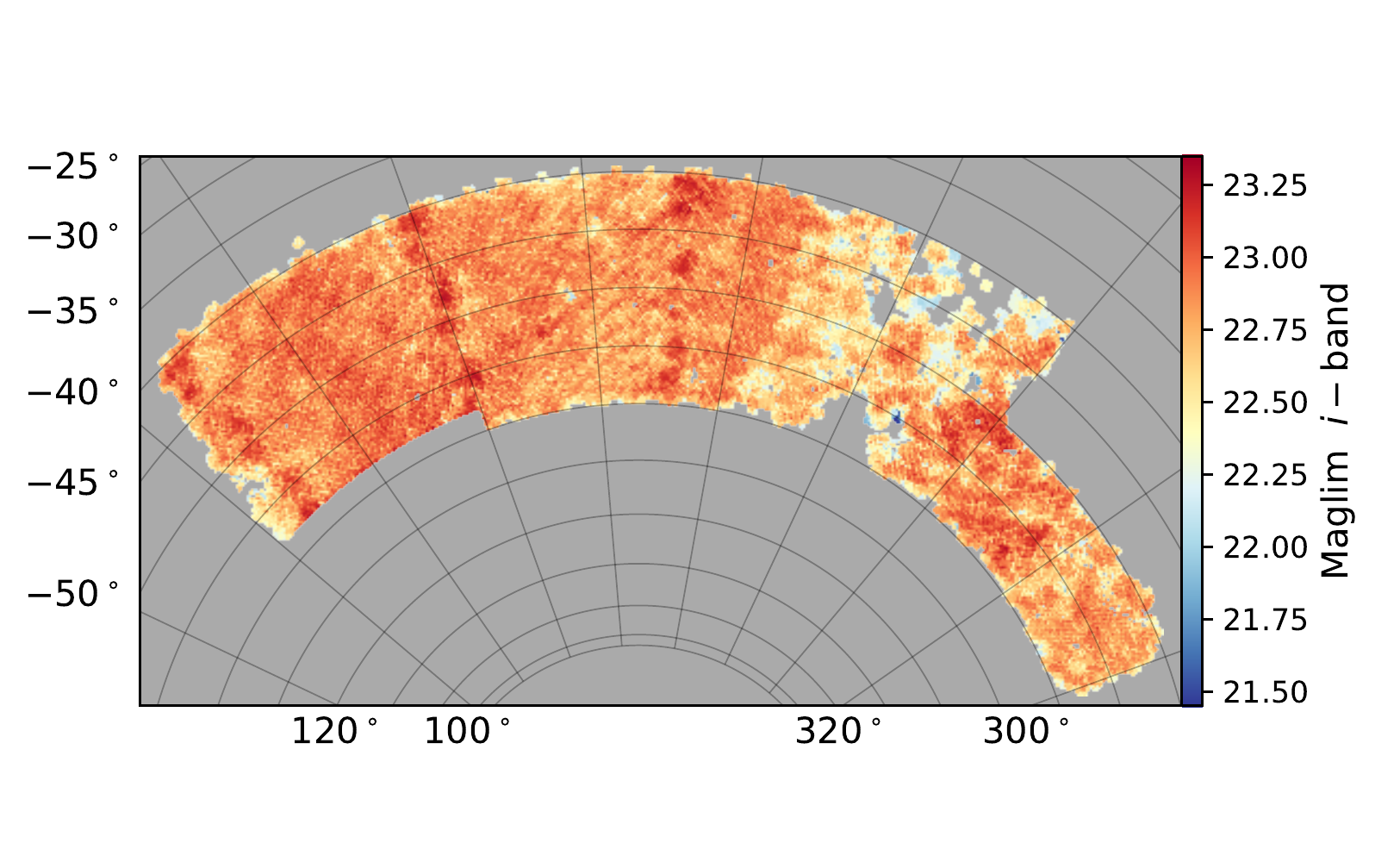}
  \\[-6mm]
  \includegraphics[scale=0.5]{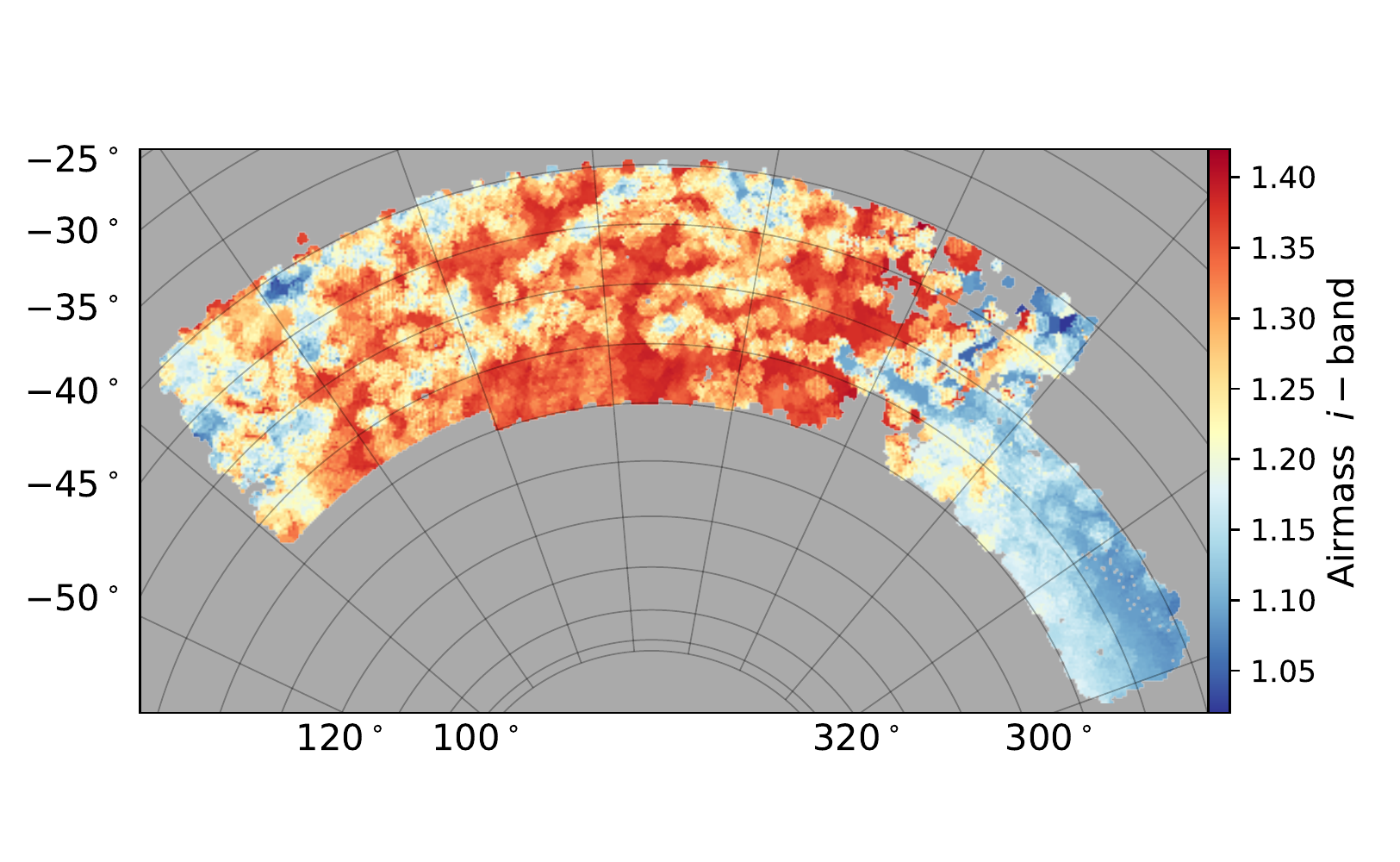}
  \includegraphics[scale=0.5]{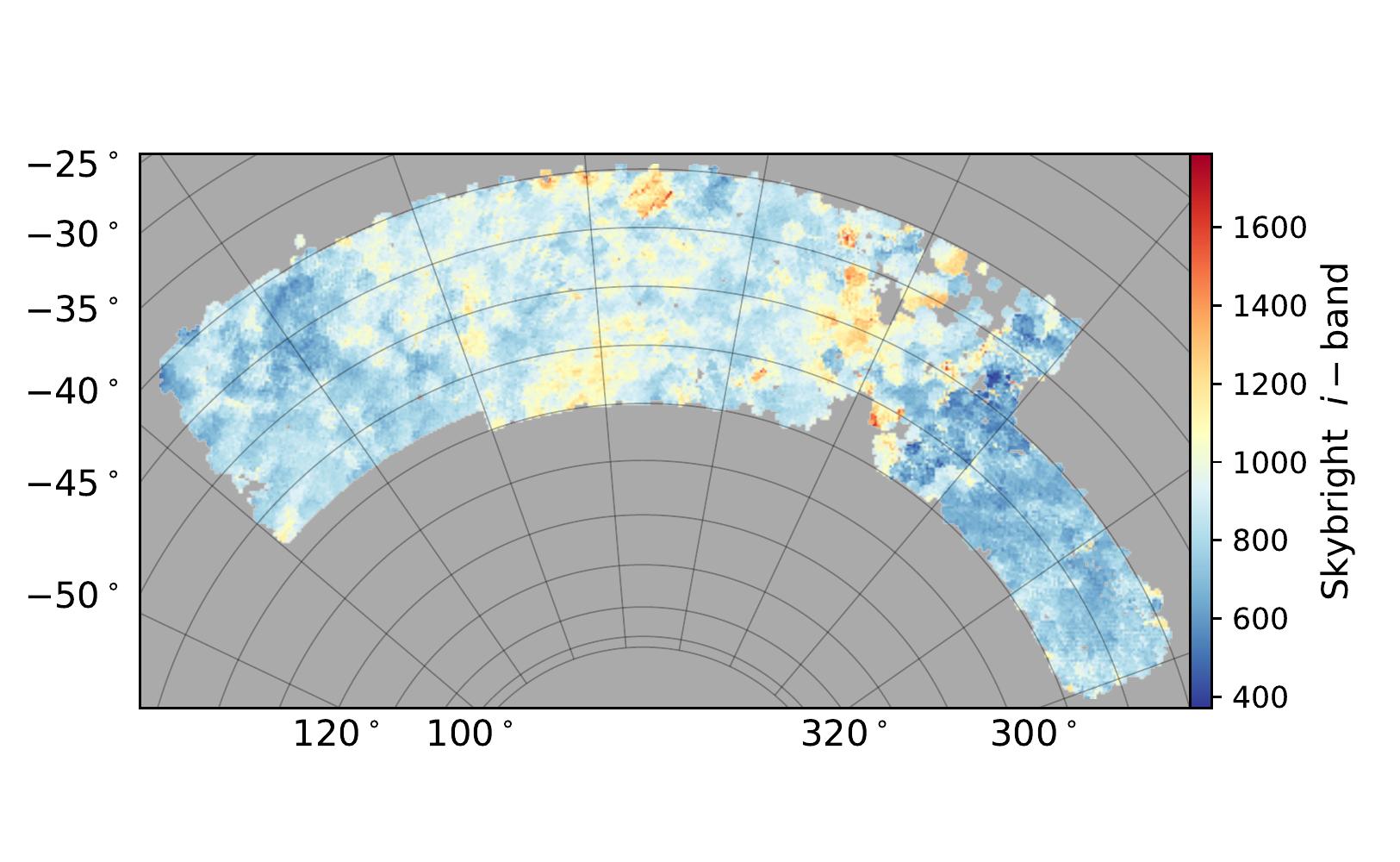}
  \\[-6
  mm]
  \includegraphics[scale=0.5]{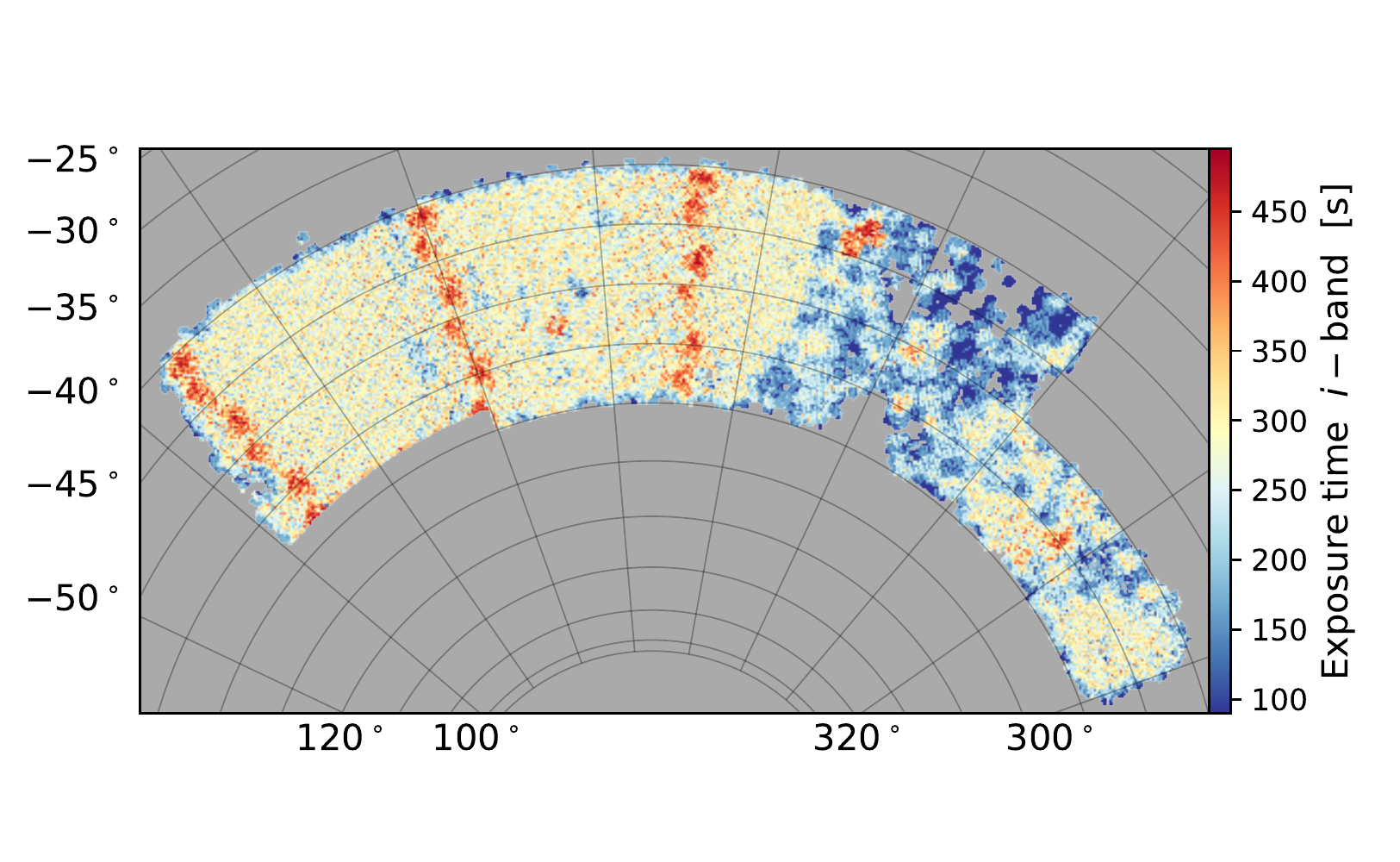}
  \includegraphics[scale=0.5]{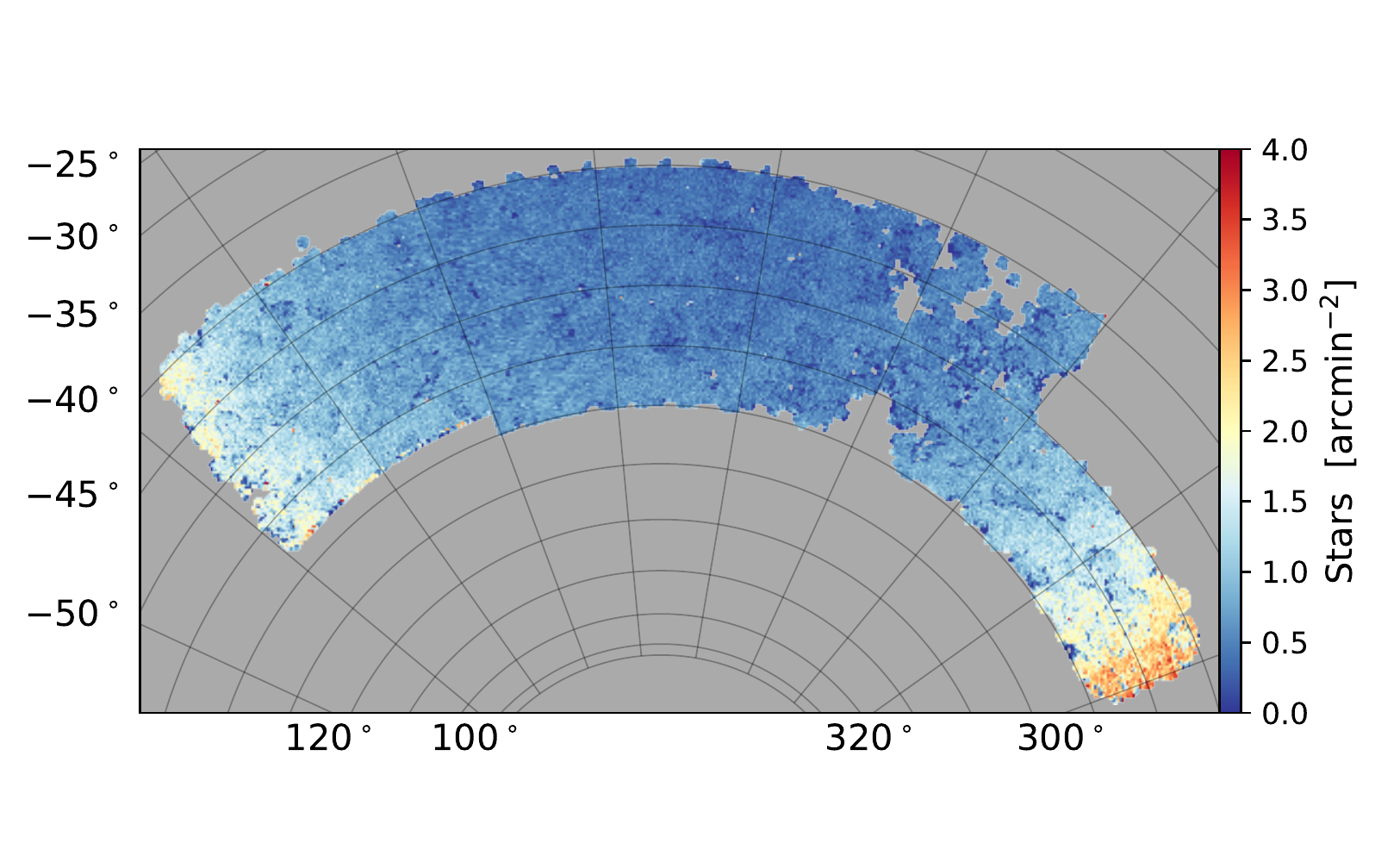}
  \caption{Maps of potential sources of systematics. Shown here for $i$-band only. Maps in other bands show fluctuations on similar scales. Each SP map is shown at $N_{\rm side} = 1024$. The stellar density map is shown at $N_{\rm side} = 512$. }
  \label{sysmaps}
\end{figure*}

In this appendix we present some examples of the survey property maps used throughout the analysis. These can be seen in Fig \ref{sysmaps}. 

\begin{figure*}
  \includegraphics[scale=0.4]{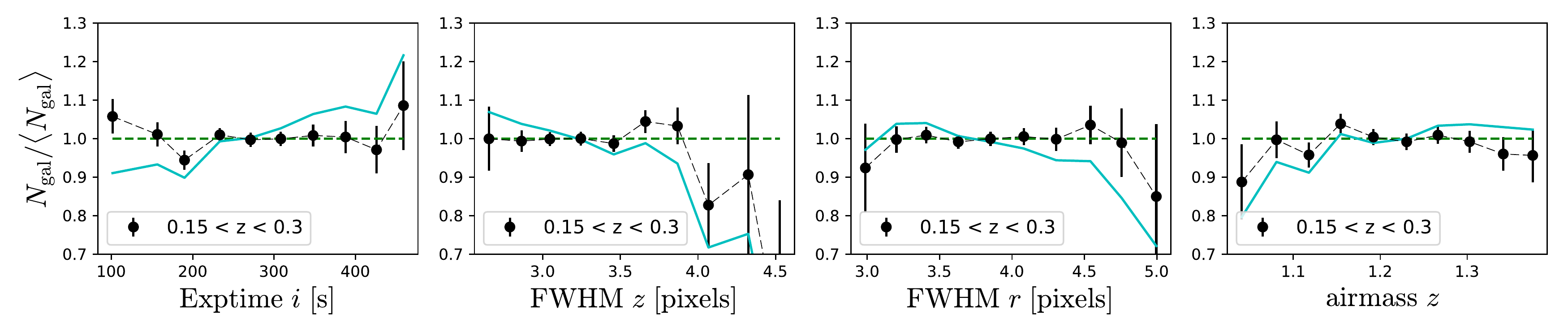}
  \includegraphics[scale=0.4]{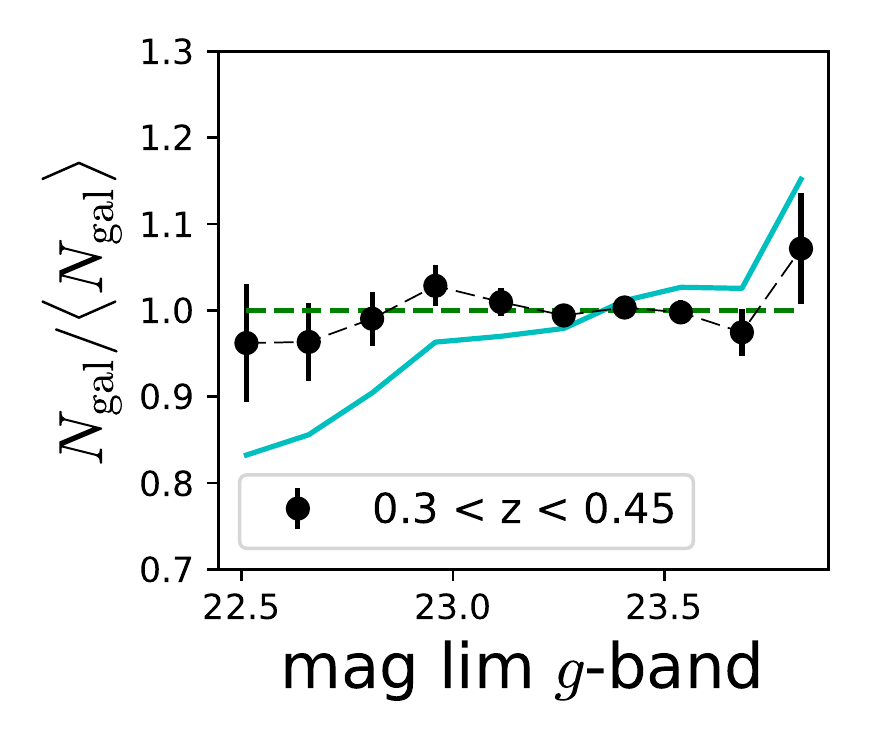}
  \hspace{0mm}
  \includegraphics[scale=0.4]{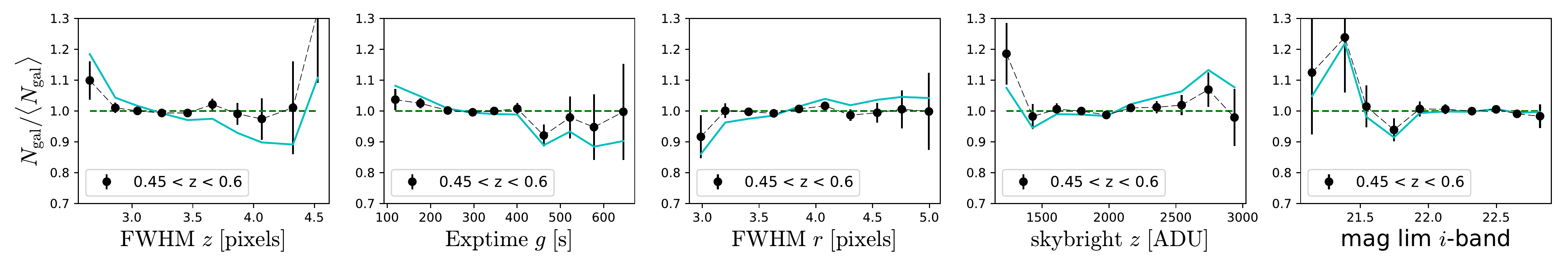}
  \includegraphics[scale=0.4]{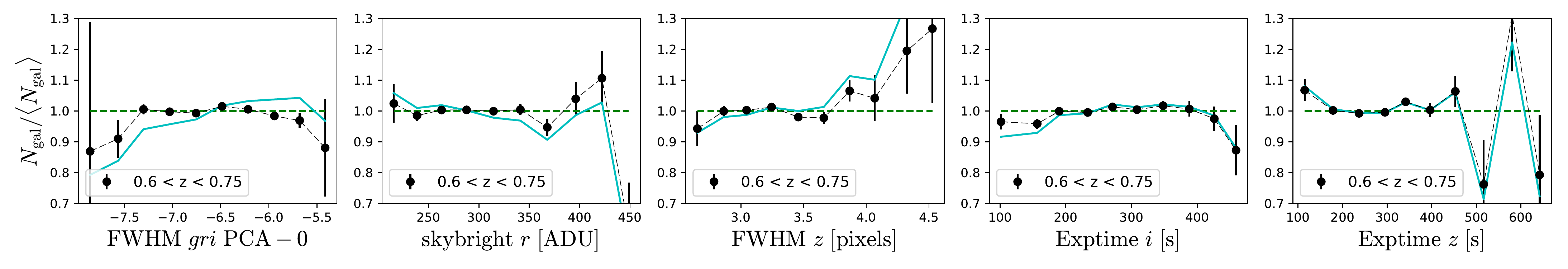}
  \includegraphics[scale=0.4]{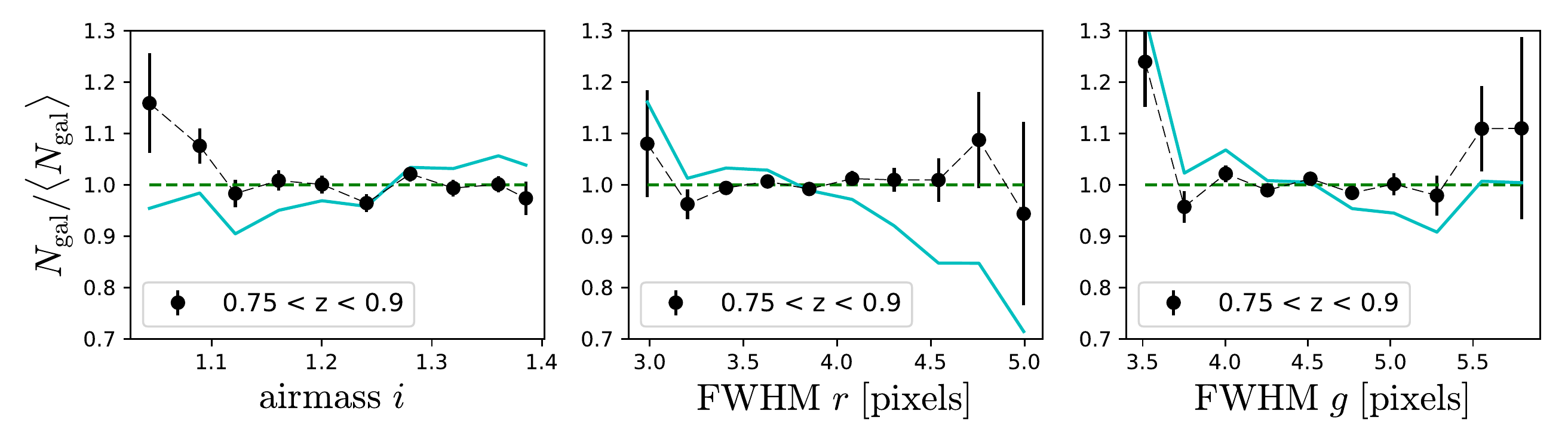}
  \caption{Galaxy number density as a function of different SP maps. We show here only the correlations with SP maps used in the $2\Delta \chi^{2}(68)$ weights calculation. The cyan line is the correlation of the sample without weights. The black points show the correlation after correction with the $2\Delta \chi^{2}(68)$ weights. The error bars were calculated by measuring the same correlation on the Gaussian mock surveys described in Section \ref{sec:covariances}. The significance of these correlations are shown in Figure \ref{null_test_plot}. }
  \label{1dsys_story}
\end{figure*}

In Fig \ref{1dsys_story} we show the correlations between the galaxy density and all the SP maps listed in Table \ref{weights_table}. 

\section{Cross correlations}
\label{appendix:cross}

\begin{figure*}
  \includegraphics[trim= 1cm 2cm 4cm 4cm, clip, scale= 0.68]{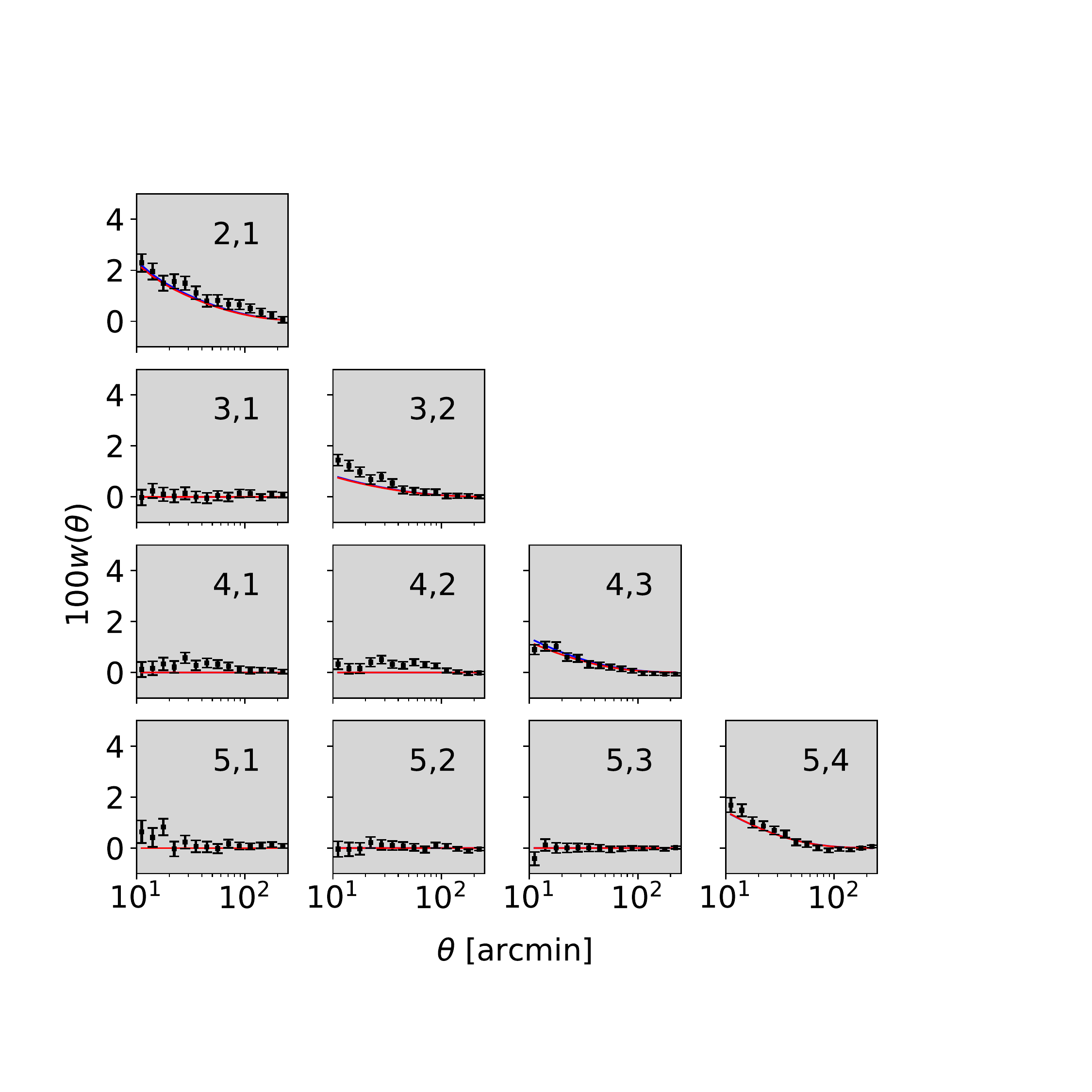}
	\caption{The two-point cross correlations between redshift bins. These measurements are expected to be non-zero, with a significance related to the degree of overlap in the $n(z)$ displayed in Fig. \ref{dndz}. The numbers in each panel correspond to the redshift bins used in the cross-correlation, Adjacent bins are shown on the diagonal. The error-bars were calculated using the log-normal mock surveys used for Y1COSMO covariance validation \citep{MPPmethodology}. The solid red curve is the best-fit model from the auto-correlation using only $w(\theta)$ at fixed cosmology. The solid blue curve (mostly under the red) is the best-fit model from the full cosmological analysis in Y1COSMO. For many of the cross-correlation panels, these predictions are indistinguishable. These measurements were not used in any of the parameter fits in this work or in Y1COSMO. }
  \label{fig:cross}
\end{figure*}

In this appendix we present the galaxy clustering signal between redshift bins, Figure \ref{fig:cross}. For these cross-correlations, we use a covariance matrix calculated from log-normal simulations described in \citep{MPPmethodology}; the square root of the diagonal of this covariance matrix yields the error-bars shown in the figure. These are the same simulations used to validate the Y1COSMO covariance matrix. 

We overplot the cross-correlation prediction both from the best fit bias values from the auto-correlations, and the best fit cosmology and bias from Y1COSMO. The cross-correlation measurements were not used in the combined probes analysis and so the robustness tests were not performed on these measurements. We present these results to demonstrate that there is a clustering signal in adjacent redshift bins (2,1), (3,2), (4,3), and (5,4) and not as a robustness test, hence we do not include a goodness-of-fit for this measurement. The amplitude of this signal is determined by the overlap in the $n(z)$ between redshift bins (see Figure \ref{dndz}). These correlations could be used in future analyses to constrain the redshift bias parameters $\Delta z^{i}$.

\end{document}